\documentclass[twocolumn]{aastex701} 

\shorttitle{Giant Planet Analogs with SPHEREx}
\shortauthors{Kiman et al.}
\received{August 9 2026}

\submitjournal{AJ}
\graphicspath{{./}{Figures/}}

\begin{document}

\title{Atmospheric Diversity of Giant Planet Analogs from 3--5\,${\rm \mu m}$ with SPHEREx}

\author[0000-0003-2102-3159]{Rocio Kiman}
\affiliation{Department of Physics, University of California, Santa Barbara, Santa Barbara, CA 93106, USA}
\email[show]{rociokiman@gmail.com}
\correspondingauthor{Rocio Kiman}

\author[0000-0003-2649-2288]{Brendan P. Bowler}
\affiliation{Department of Physics, University of California, Santa Barbara, Santa Barbara, CA 93106, USA}
\email{bpbowler@ucsb.edu}

\author[0000-0002-6618-1137]{Jerry W. Xuan}
\altaffiliation{51 Pegasi b Fellow}
\affiliation{Department of Earth, Planetary, and Space Sciences, University of California, Los Angeles, CA 90095, USA}
\email{jerryxuan@g.ucla.edu}

\author[0009-0003-5850-464X]{Narisara Mayer}
\affiliation{Department of Physics, University of California, Santa Barbara, Santa Barbara, CA 93106, USA}
\email{nmayer@ucsb.edu}

\author[0000-0002-9884-9584]{Claire Finley}
\affiliation{Department of Physics, University of California, Santa Barbara, Santa Barbara, CA 93106, USA}
\email{clairefinley@ucsb.edu}

\begin{abstract}

The emergent spectra of giant planets peak at thermal infrared wavelengths (3--5\,${\rm \mu m}$) and possess atmospheric features in this regime that trace carbon chemistry, vertical mixing, clouds, and composition.  However, this spectral region is challenging to access from the ground, and with JWST generally requires an independent pointing for each target. SPHEREx is providing all-sky spectra of millions of sources from 0.75--5\,${\rm\mu m}$, enabling both individual and population-level studies of isolated and widely bound substellar objects. In this work we present a SPHEREx spectrophotometric atlas of 94 young ($\lesssim$300~Myr) low-gravity L and T dwarfs to study their atmospheric diversity in the relatively unexplored wavelength range from 3--5\,${\rm \mu m}$. These include free-floating planetary-mass objects in a range of ages, masses, and temperatures regularly probed with JWST and which will be accessible with upcoming instruments such as Keck/SCALES and ELT/METIS. We compute synthetic JWST/NIRCam and MKO photometry and colors across 22 filters, and examine trends in physically-motivated sets of colors, absolute magnitudes, and spectral types.
We find that the lowest-gravity objects generally have the reddest 3--5\,${\rm \mu m}$ colors and tend to be brighter than field L dwarfs for a given spectral type, extending previous patterns originally established at near-infrared wavelengths.
Compared to the low-gravity sample of free-floating objects, some giant planets closely follow the low-gravity locus, while others show larger scatter and more extreme colors, pointing to differences in cloud properties, temperature, metallicity, or vertical mixing.
The SPHEREx data extraction code developed in this work is made available together with the extracted spectra and synthetic photometry.

\end{abstract}



\section{Introduction}

Directly imaged giant planets and their substellar analogs---low-gravity brown dwarfs---offer a unique window into the physics of giant planet atmospheres. The spectra of these objects contain information about their composition, atmospheric dynamics, and bulk physical properties, which can in turn be used to constrain how planets form and evolve \citep[e.g.,][]{Marley2015,Bowler2016,Biller2017}. For young, low-mass companions, their low surface gravities allow clouds to persist high in the atmosphere, producing the red near-infrared colors that distinguish them from older field brown dwarfs of similar effective temperature \citep{2016ApJ...833...96L,2016ApJS..225...10F}. This connection between youth, low gravity, cloudy atmospheres, and red photometric colors has been established through systematic studies of directly imaged planets such as HR 8799 bcde \citep{Marois2008,Marois2010,Bowler2010,Barman2011,Marley2012} and brown dwarf analogs such as VHS~1256--1257~b \citep{Gauza2015,Miles2023}.

The near-infrared (1--$2.5\,{\rm\mu m}$) and thermal-infrared (3--$5\,{\rm\mu m}$) wavelength regimes are particularly powerful for characterizing these objects. In these ranges, absorption bands from ${\rm H_2O}$  at $0.9\, {\rm \mu m}$, $1.2\, {\rm \mu m}$, $1.4\, {\rm \mu m}$, $1.9\, {\rm \mu m}$ and $2.7\, {\rm \mu m}$, CO at $2.29\,{\rm \mu m}$ and $4.6\,{\rm \mu m}$, ${\rm CH_4}$ at $1.17\,{\rm \mu m}$, $1.7\,{\rm \mu m}$, $2.3\,{\rm \mu m}$ and $3.3\,{\rm \mu m}$, and ${\rm CO_2}$ at $4.3\,{\rm \mu m}$ \citep{Cushing2005,Yamamura2010} constrain atomic abundance ratios such as C/O, C/H, and O/H, and reveal the degree of disequilibrium chemistry driven by atmospheric vertical mixing \citep[e.g.,][]{Konopacky2013, Barman2015, Molliere2020, Xuan2024}. This 1--$5\,{\rm\mu m}$ region is also sensitive to collision-induced absorption from ${\rm H_2}$, as well as bulk metallicity \citep[e.g.,][]{Leggett2010}.

In the near-infrared, gravity-sensitive atomic and molecular spectral features have long been used to identify and classify low-gravity objects \citep{Cruz2009,2013ApJ...772...79A}. The characteristic triangular $H$-band peak and reduced alkali line strengths are especially prominent signatures of youth and low gravity for giant planets and brown dwarfs \citep{Lucas2001,Gorlova2003,Kirkpatrick2006,2013ApJ...772...79A,Bowler2014}. 
For example, the widely-separated substellar companion VHS~1256--1257~b shares photometric colors and spectroscopic features with HR 8799 c, d, and e; features of youth and low surface gravity; suppressed methane absorption; and prominent CO absorption indicative of strong disequilibrium chemistry \citep{Miles2018,Miles2023}.
Obtaining high-quality near-infrared spectra for large samples of low-gravity L and T dwarfs has historically required significant observational resources. These objects are faint, and existing spectroscopic surveys have assembled samples with variable completeness and coverage \citep[e.g.,][]{2015ApJS..219...33G,Manjavacas2020,Piscarreta2024}.

At longer wavelengths, the thermal infrared regime provides a probe of molecular species inaccessible to shorter wavelengths.  Warm instruments, telescope optics, and Earth's atmosphere make ground-based photometry and spectroscopy challenging \citep[e.g.,][]{Golimowski2004,Skemer2014,Morley2018}.
The launch of the James Webb Space Telescope \citep[JWST,][]{2023PASP..135f8001G} has transformed our ability to study giant planet and brown dwarf atmospheres across the full $1–5\,{\rm \mu m}$ wavelength range. Early results from JWST/NIRSpec and NIRCam have provided unprecedented detections of molecules including ${\rm H_2O}$, CO, ${\rm CH_4}$, ${\rm CO_2}$, ${\rm H_2S}$, and NH$_3$ in directly imaged planets and their analogs \citep[e.g.][]{Miles2023,Franson2024,Balmer2025, Malin2025,Ruffio2026,Xuan2026,Kiman2026}. In particular, ${\rm CO_2}$ at $4.3\,{\rm \mu m}$ has emerged as a sensitive tracer of atmospheric metallicity \citep{Balmer2025,Balmer2026}, readily accessible from space but difficult to access from the ground as a result of bright telescope and sky emission, and strong telluric absorption from atmospheric ${\rm CO_2}$. However, conducting large-scale spectroscopic surveys of long-period planets and individual field L and T dwarfs with JWST is both expensive and time-intensive.

Complementing JWST's targeted high-resolution view, the Spectro-Photometer for the History of the Universe, Epoch of Reionization, and Ices Explorer \citep[SPHEREx,][]{Bock2026} mission is providing all-sky low-resolution spectrophotometry covering the $0.75–5\,{\rm \mu m}$ wavelength range for nearly 10 million sources to 22 mag by the end of the mission. SPHEREx is already revolutionizing the field of ultracool dwarfs \citep{Rustamkulov2026,Tu2026,Gagne2026,Brooks2026}
and offers an unprecedented opportunity to study low-gravity brown dwarfs and giant planet analogs at population scales, without the observational overhead required for targeted spectroscopy. A well-characterized sample of known low-gravity objects with SPHEREx spectra can serve both as a calibration set for spectral indices sensitive to surface gravity and youth, and as a reference for interpreting JWST observations of directly imaged planets and their atmosphere in this same wavelength range.

In this paper, we present a SPHEREx spectrophotometric atlas of 94 known low-gravity L and T dwarfs and use these to explore trends in absolute magnitude, color, and spectral type---with an emphasis on the less-explored 3--$5\,{\rm \mu m}$ range. By comparing with expectations from atmospheric models, we connect these relationships to physical changes in cloud properties, surface gravity, and vertical mixing.
In Section~\ref{sec:extraction} we describe the SPHEREx mission and the code we developed to extract the spectra, together with the tests we carried out using existing data in the literature to validate our method. In Section~\ref{sec:lowgravitysequence} we describe our sample of low-gravity L and T dwarfs and present the SPHEREx spectral sequence across the L and T types. Using this sample, in Section~\ref{sec:jwstphotometry} we calculate JWST/NIRCam photometry and compare the color-color and color-magnitude positions of low-gravity and field L and T dwarfs with known giant exoplanets. Finally, in Section~\ref{sec:conclusions} we summarize our conclusions.

\section{Extraction of SPHERE\lowercase{x} Spectrophotometry} 
\label{sec:extraction}

\subsection{The SPHEREx mission}

SPHEREx is a two-year mission conducting the first all-sky spectral survey in the near-infrared. Launched on March 11, 2025, the satellite surveys the entire sky every six months across 102 spectral bands spanning 0.75 to 5.0\,${\rm \mu m}$.
The instrument achieves this coverage using six linear variable filters (LVFs), each producing a wavelength gradient across its detector. Every detector is divided into 17 standard spectral channels, yielding the full 102-band coverage. The six filters cover different wavelength ranges and spectral resolutions: Band 1: $\lambda = 0.75 - 1.09\,{\rm \mu m}$ with $R=41$, where $R = \lambda / \Delta \lambda$; Band 2: $\lambda = 1.10 - 1.62\,{\rm \mu m}$ with $R=41$; Band 3: $\lambda = 1.63 - 2.41\,{\rm \mu m}$ with $R=41$; Band 4: $\lambda = 2.42 - 3.82\,{\rm \mu m}$ with $R=35$; Band 5: $\lambda = 3.83 - 4.41\,{\rm \mu m}$ with $R=110$; Band 6: $\lambda = 4.42 - 5.00\,{\rm \mu m}$ with $R=130$. Each pixel has a size of $6\farcs 15$ on the sky. The SPHEREx team is making the data public as it is taken by the telescope with different levels of processing. In this study, we worked directly with the Level 2 data products \citep{spherex} from the SPHEREx pipeline \citep{Akeson2025}, which provides photometrically and astrometrically calibrated spectral images and data cubes.

\begin{figure*}[ht!]
\begin{center}
\includegraphics[width=\linewidth]{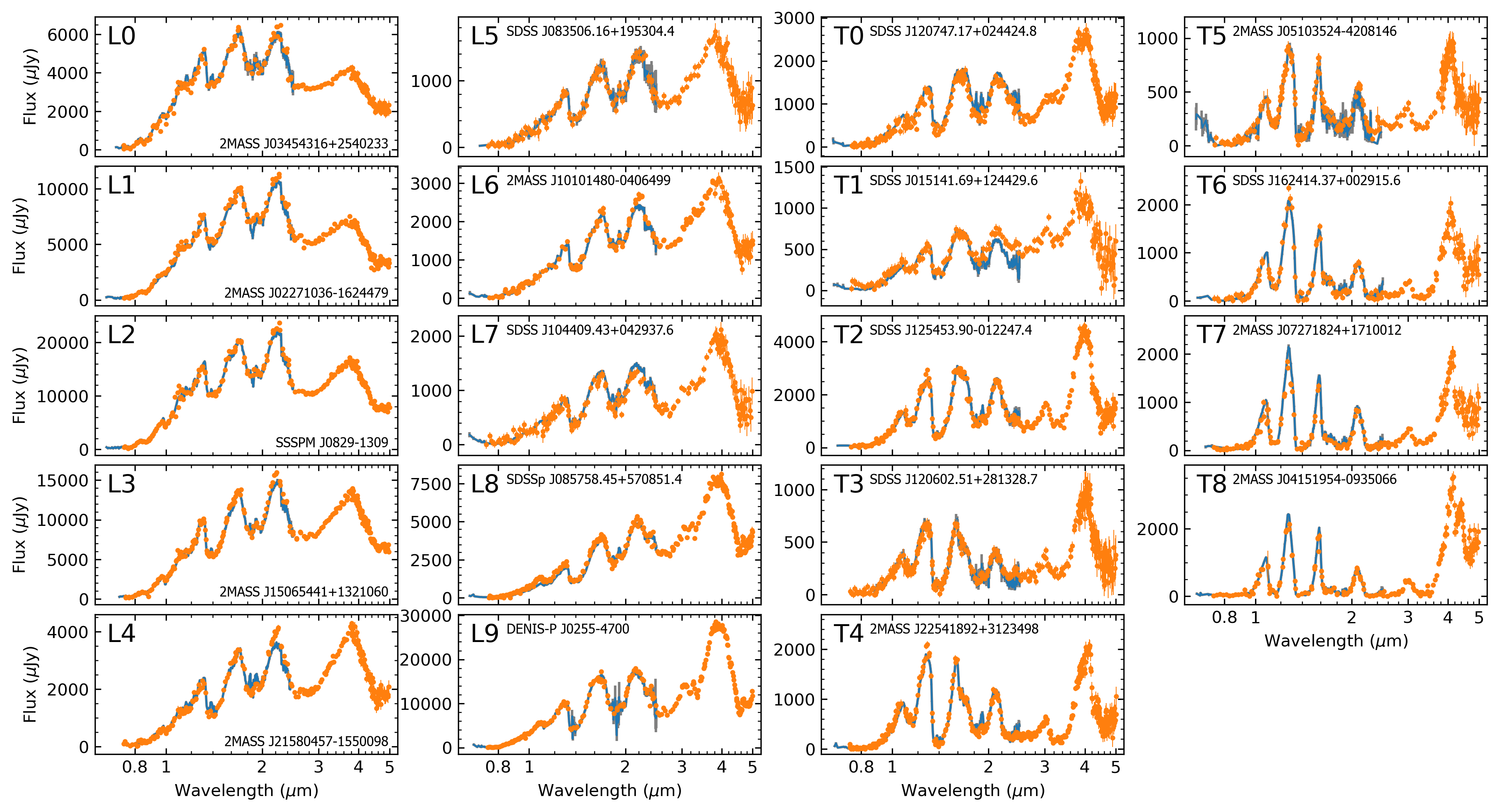}
\caption{Sample of field L and T dwarf standards to test our extraction code, \texttt{spherexminer}. We show the SPHEREx extraction in orange, compared to 0.85--2.45\,${\rm \mu m}$ IRTF/SpeX prism spectra flux calibrated to 2MASS photometry in blue. In general the SPHEREx spectra are in excellent agreement with ground-based near-infrared spectra, both in terms of their absolute flux levels and overall spectral shape.  Note that the SPHEREx data have not themselves separately been flux-calibrated to photometry.} 
\label{fig:standards}
\end{center}
\end{figure*}

\subsection{Extraction Code: \texttt{spherexminer}}

We developed \texttt{spherexminer}\footnote{\url{https://github.com/rkiman/spherexminer}}, a Python pipeline for extracting spectra of individual sources from the Level 2 images available on IRSA, following the IRSA tutorial\footnote{\url{https://caltech-ipac.github.io/irsa-tutorials/spherex-intro/}}. The pipeline proceeds as follows. Using the \texttt{IRSA} module from \texttt{astroquery} \citep{2019AJ....157...98G}, we identify the images closest to the position of each target based on input coordinates and appropriate corrections for proper motion. Because objects are observed multiple times at different epochs, we determine each source's position in every image independently via proper motion propagation—an important step for fast-moving objects. We then perform aperture photometry on each image using \texttt{photutils} \citep{bradley_2026_19636730}, applying the quality flags recommended in the SPHEREx Explanatory Supplement to reject contaminated background regions and bad pixels.\footnote{Specifically, we removed pixels with the following background flags: OVERFLOW, SUR\_ERROR, NONFUNC, MISSING\_DATA, HOT, COLD, NONLINEAR, PERSIST, OUTLIER, SOURCE, TRANSIENT, BLOOM, SNOWBALL, HALO, SATELLITE\_HALO. To calculate the aperture flux, we removed the following flags: SUR\_ERROR, NONFUNC, MISSING\_DATA, HOT, COLD, NONLINEAR, PERSIST, BLOOM, SNOWBALL, HALO, SATELLITE\_HALO.} 
For each image, the aperture radius is set to enclose $95\%$ encircled energy, based on a position-dependent PSF model evaluated at each object's location. We use the exposure-averaged, position-dependent PSF models provided by the SPHEREx Science Data Center, which are derived from calibration-star observations and account for the effective instrumental response, including telescope pointing jitter. Half of the targets included in this work required a $90\%$ or $98\%$ encircled energy to improve agreement with literature magnitudes (2MASS $J$, $H$, $K_{\rm s}$, WISE W1 and W2, Spitzer ch1 and ch2, and MKO $M'$ and $L'$). 
For each aperture we apply the corresponding aperture correction.
The source flux is computed as the background-subtracted sum within the aperture, with local background estimated from the median of an annulus scaled to the aperture area, and the corresponding uncertainty propagated from the per-pixel error array.
The wavelength at each object's image position is computed by transforming its pixel coordinates through the image's spectral World Coordinate System solution, using the linear distortion terms to obtain the corresponding wavelength. 
Thus from each image we obtain a wavelength, flux, and flux uncertainty; the total number of measurements per object depends on how many survey passes cover that position.

The large SPHEREx pixel scale makes aperture photometry susceptible to contamination from nearby sources. We therefore visually inspect every extracted spectrum and discard any object with a neighbor within $\sim 10''$. 
Several additional refinements are applied to the extracted spectra.
Isolated outlier points corresponding to bad pixels and contamination from nearby stars---typically fewer than 10 flux density measurements per spectrum---are removed. Finally, we discard the highest-uncertainty points ($>50\%$) near $1.10\,{\rm \mu m}$, where the Band 1–2 transition degrades the extraction reliability.

\subsection{Test of \texttt{spherexminer} using brown dwarf standards}
\label{subsec:standards}

To validate the pipeline-processed flux calibration across the entire 0.75--$5\,{\rm \mu m}$ extracted spectra, we assemble a sample of standard L and T dwarfs drawn primarily from the curated list of \citet{Lueber2022}. 
These objects represent traditionally defined standard L and T dwarfs \citep[e.g.,][]{2006ApJ...637.1067B,2010ApJS..190..100K}---single, bright, and well-characterized objects that are representative of their spectral subclass.  
We replace several objects from the original list where nearby contaminants do not allow a clean extraction or where comparison spectra could not be obtained from the literature. The new objects for this analysis are 2MASS J02271036-1624479 (L1), SDSS J104409.43+042937.6 (L7), SDSSp J085758.45+570851.4 (L8), and 2MASS J05103524-4208146 (T5). None of these objects show evidence of binarity or other anomalous properties in the literature, and their ($J-K_{\rm s}$) colors are consistent with  brown dwarfs of the same subtype from the 20\,pc field sample of \citet{2021ApJS..253....7K}.
All standard objects have ground-based IRTF/SpeX \citep{Rayner2003} spectra drawn from the SpeX Prism Libraries \citep{Burgasser2004,Burgasser2006,2006ApJ...637.1067B,Burgasser2006c,Burgasser2007,Burgasser2007b,Burgasser2008,Burgasser2010,Chiu2006,Looper2007,Reid2006}, covering 0.85--$2.45\,{\rm \mu m}$ at $\lambda/\Delta\lambda \sim$ 85--300. For comparison, we flux-calibrate the IRTF/SpeX spectra to 2MASS photometry, with one exception: for the T8 dwarf 2MASS J04151954-0935066, we use the UKIRT/WFCAM $J$, $H$ and $K$ magnitudes from \citet{Vrba2026}, which provided more precise measurements compared to 2MASS, owing to the object's faintness ($K$=16~mag). 
Figure~\ref{fig:standards} compares the extracted SPHEREx data to the IRTF/SpeX spectra for the full standard sample.

The spectra are generally in excellent agreement.  In some cases, there are signs that the individual heights of some features appear to disagree slightly (at the $\sim$ 10$\%$ level)---for example, the $K$-band peak of the L7 standard SDSS J104409.43+042937.6.
Similarly, there appear to be subtle chromatic differences for the T1 standard SDSS J015141.69+124429.6. This may be caused by intrinsic gray or wavelength-dependent variability, or systematics in either the SPHEREx spectra or the SpeX spectra (for example, chromatic changes caused by telluric correction or differential atmospheric refraction).
At wavelengths beyond $2.4\,{\rm \mu m}$, the rich and diverse set of molecular absorption features is clearly evident.

\section{SPHERE\lowercase{x} Low-gravity L and T dwarfs sequence}
\label{sec:lowgravitysequence}

We use SPHEREx data to study the diversity of low-gravity L/T dwarfs. Starting from the UltracoolSheet \citep{UltracoolSheet}, we select all objects flagged as low-gravity, excluding known binaries, yielding a sample of 154 objects. This includes objects that are classified as low-gravity using optical and/or infrared indicators. 
As these objects have features of low-gravity and thus are still contracting, they are likely younger than a few hundred Myr \citep{Burrows2001,2016ApJ...833...96L}. 

The sample is dominated by early L dwarfs, with a smaller number of mid-to-late L dwarfs and two T dwarfs. Surface gravities are characterized using the optical Greek-letter classification (\citealt{Cruz2009}; $\beta$ and $\gamma$) and/or the near-infrared gravity classification of \citet[INT-G and VL-G, with FLD-G denoting normal field gravity]{2013ApJ...772...79A}. Although developed independently, both schemes were designed to identify young, low-surface-gravity ultracool dwarfs.  By calibrating these gravity classifications against ultracool dwarfs with independently determined ages, primarily members of nearby young moving groups and star-forming regions, these schemes have been shown to correspond to approximate age ranges of $\lesssim$150 Myr for VL-G ($\gamma$) objects and $\sim$50–300 Myr for INT-G ($\beta$) objects \citep[e.g.,][]{2013ApJ...772...79A}. 
These classifications are based on the strength of gravity-sensitive spectral features—such as VO and FeH absorption bands and alkali lines (Na I, K I), which weaken with lower surface gravity, and the shape of the $H$-band continuum, which becomes more triangular in low-gravity objects due to reduced collision-induced ${\rm H_2}$ absorption and weaker FeH.  

These gravity classifications are not available for every object in both classification schemes. Thirteen objects in our sample are classified only as low gravity (LG) in the near-infrared literature with no corresponding optical gravity classification. In these cases, the original studies identified the objects as low gravity based on their spectral features but did not assign a more specific VL-G or INT-G designation.
All of the 154 objects in our sample have a near infrared gravity classification, while only 48 have an optical classification.
The optical and infrared classifications are consistent for 34 ($71\%$) of these objects: 28 VL-G/$\gamma$ objects and 6 INT-G/$\beta$. In the rest of our work we use the near-infrared classification for simplification.

Reliable SPHEREx spectra could be extracted for 94 of the low-gravity L and T dwarfs, which are shown in Figures~\ref{fig:spectra_comparison0} and \ref{fig:spectra_comparison1}. The rest of the objects had a nearby star which contaminated the spectra, and were discarded.
Spectral extraction and cleaning follow the same procedure described in Section~\ref{sec:extraction}. As an additional quality check, we compare synthetic 2MASS ($J$, $H$, and $K_{\rm s}$) and Spitzer (ch1) photometry against the measured values, finding agreement within $1$–$2\sigma$ and a scatter of $\sim 0.1$\,mag. The code to estimate synthetic photometry is based on the work of \citet{Suarez2021}.

Some spectra from Figures~\ref{fig:spectra_comparison0} and \ref{fig:spectra_comparison1} exhibit artifacts arising from the long-baseline nature of the SPHEREx scanning strategy. In particular, a handful of objects appear to have multiple distinct spectral traces, which can be produced when that position is revisited at separate epochs; if the object is variable, the resulting spectra will differ \citep{2020SPIE11443E..0IC}. 

Low-gravity objects are systematically redder than the corresponding field standards at longer wavelengths, as expected given the normalization over the $1.2$–$1.5\,{\rm \mu m}$ range. The severity of the deviation varies from object to object, but is most pronounced for late-L dwarfs and diminishes for T dwarfs, consistent with cloud decks sinking to greater atmospheric depths across the L/T transition.
It is clear that the same gravity-sensitive spectral behavior that has long been recognized at near-infrared wavelengths extends to longer wavelengths. The reddest objects in the sample, such as TWA 41 and 2MASS J21140802--2251358, depart dramatically from the field template in the 3--5 $\mu$m region. Most mid- to late-L dwarfs in the sample peak near 4 $\mu$m (in $f_{\nu}$), marking this the turnover point of their spectral energy distribution.

Several physical mechanisms have been proposed to explain this enhanced reddening, including low surface gravity \citep{2016ApJS..225...10F,2016ApJ...833...96L}, optically thick dust clouds \citep{Marocco2014}, super-solar metallicity \citep{Looper2008}, and viewing inclination \citep{Vos2020}. Many of these parameters are expected to be correlated, and there is a clear diversity of 3--5 $\mu$m spectral behavior among objects with the same spectral type and gravity classification.

\begin{figure*}[ht!]
\begin{center}
\includegraphics[width=0.9\linewidth]{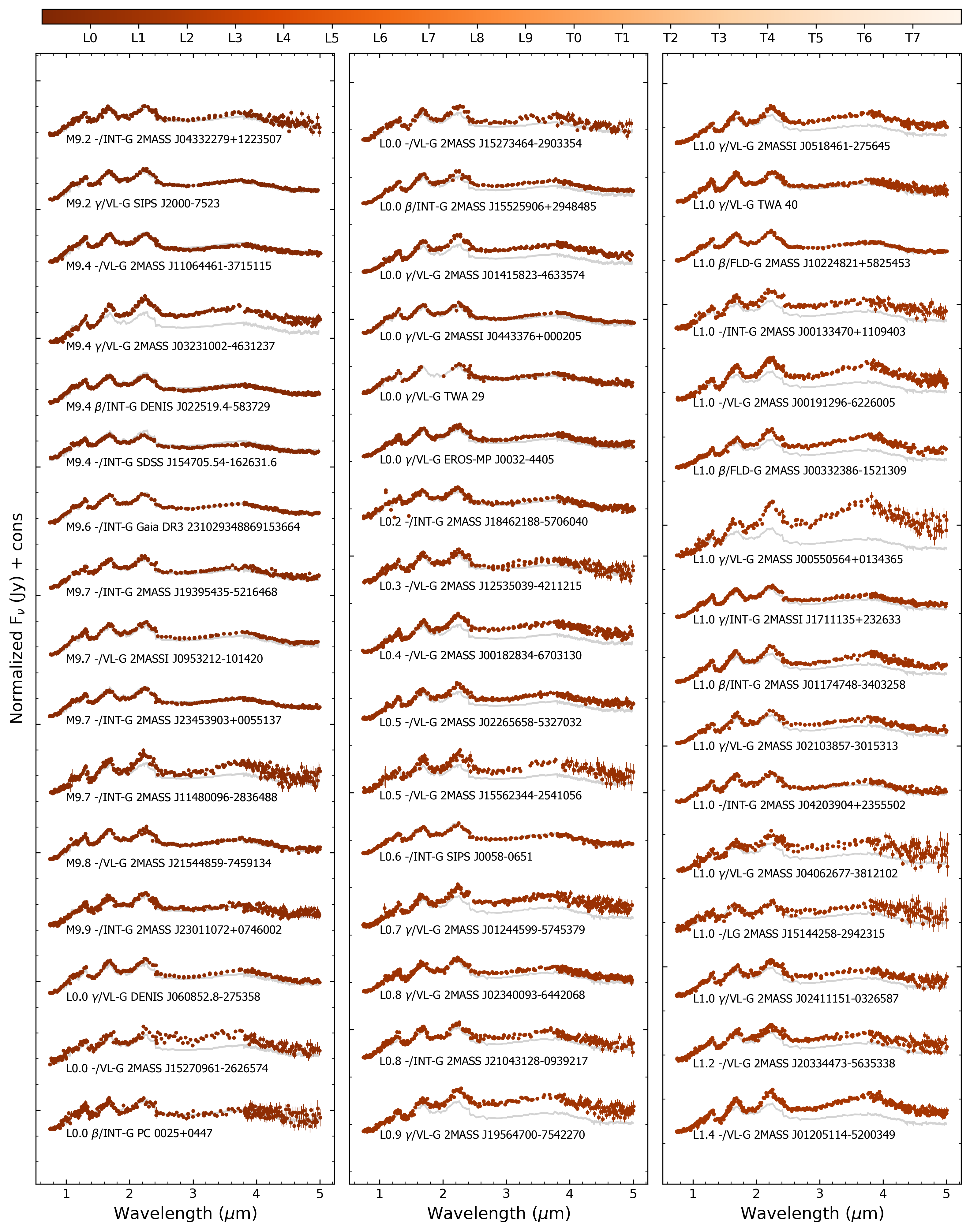}
\caption{SPHEREx spectra extracted with our code \texttt{spherexminer} for low-gravity objects in the UltracoolSheet. Each spectrum is color-coded according to infrared spectral type. We also show for each spectrum the closest corresponding field standard in gray; for these objects, the width of the line represents the uncertainty across the spectrum. Both low-gravity and standard spectra are normalized between $1.2$–$1.5\,{\rm \mu m}$. In most cases, the comparison to the standard field star reflects directly the gravity classification. Each label below the spectra indicates the infrared spectral type, optical/infrared gravity classification and name of the object. All the spectra can be found in Zenodo \citep{kiman_2026_21856313}\footnote{\url{https://zenodo.org/records/21856313}}.} 
\label{fig:spectra_comparison0}
\end{center}
\end{figure*}

\begin{figure*}[ht!]
\begin{center}
\includegraphics[width=0.9\linewidth]{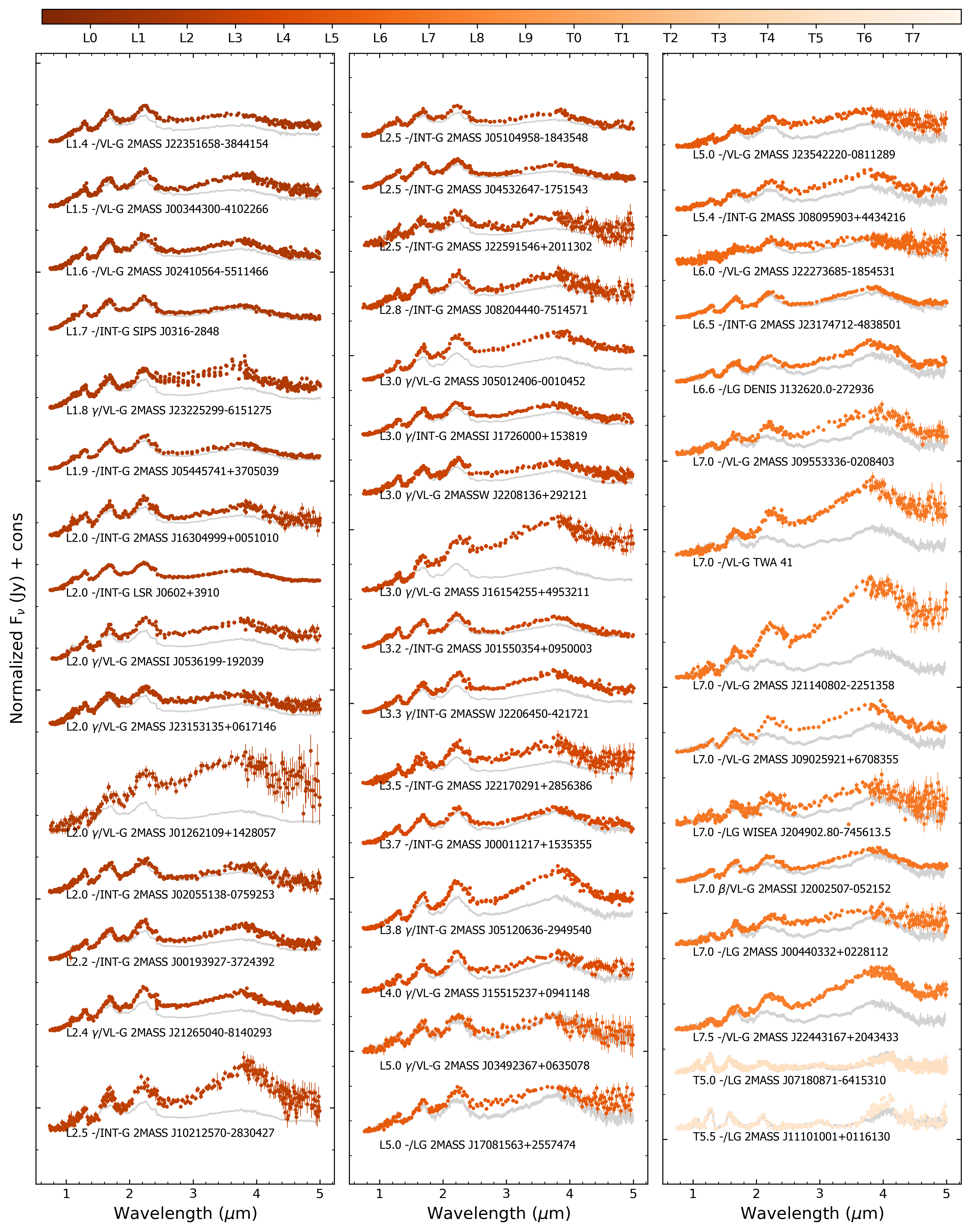}
\caption{Continuation of   Figure~\ref{fig:spectra_comparison0}.} 
\label{fig:spectra_comparison1}
\end{center}
\end{figure*}

\section{JWST/NIRCam Synthetic Photometry and Colors}
\label{sec:jwstphotometry} 

JWST/NIRCam is one of the primary instruments for imaging and characterizing directly imaged giant planets, making it important to establish empirical color sequences for young substellar objects observed in these bandpasses. Young, low-gravity brown dwarfs provide a large sample of analogs spanning a broad range of spectral types and atmospheric properties, enabling a systematic exploration of how NIRCam colors vary with gravity and temperature.  Here, we use JWST/NIRCam filters to study the diversity of low-gravity L and T dwarfs, using the field dwarf sample described in Section~\ref{subsec:standards} as a reference. Because young, low-gravity brown dwarfs serve as analogs to directly imaged giant planets, we also compare our sample to imaged giant planets with existing JWST observations.

Below, we describe the JWST/NIRCam filters selected for our analysis and use them to examine the dependence of color and absolute magnitude on spectral type and gravity classification. 
We also explore the behavior of field objects, low-gravity brown dwarfs, isolated planets, and directly imaged giant planets in representative color-color and color-magnitude diagrams.

\subsection{JWST/NIRCam filters}

\begin{figure*}[ht!]
\begin{center}
\includegraphics[width=\linewidth]{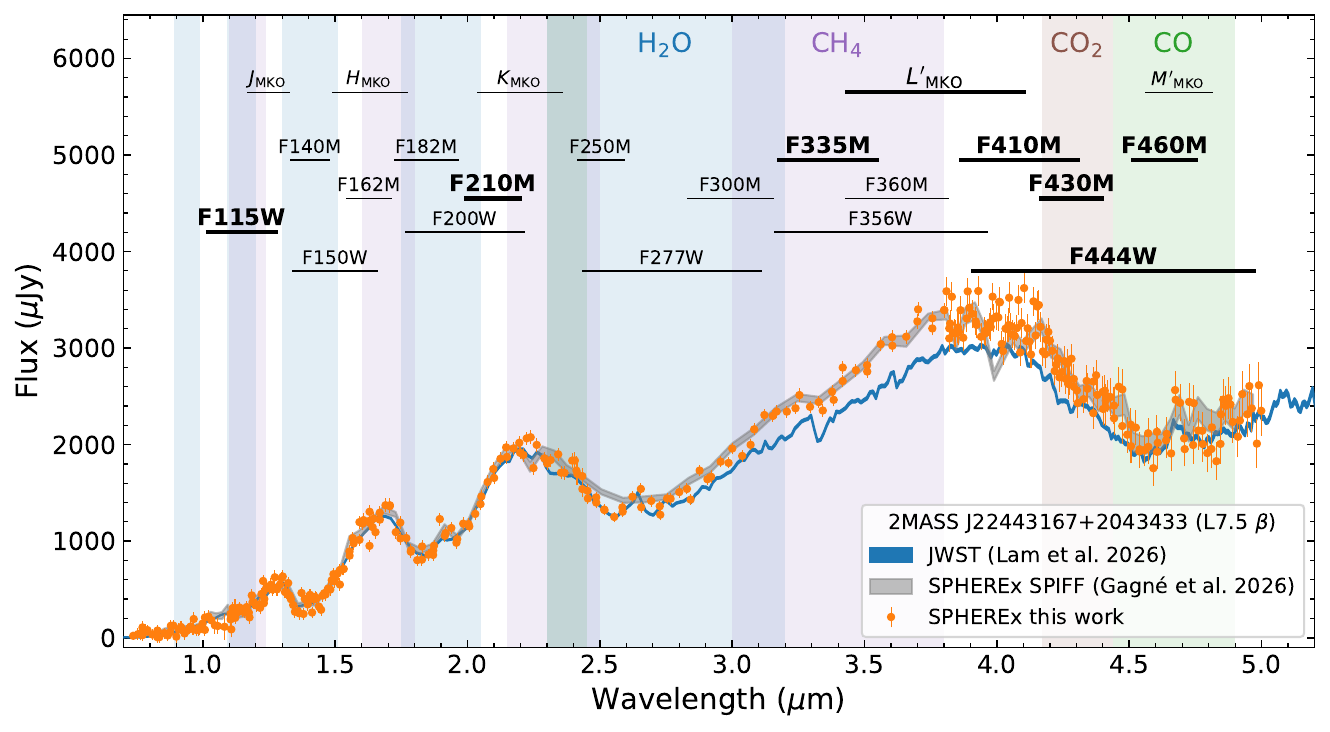}
\caption{JWST/NIRCam filters, plus MKO $J$, $H$, $K$, $L'$ and $M'$, with the filters selected for this paper highlighted in bold. The NIRSpec Prism spectrum for 2MASS J22443167+2043433 (L7.5 $\beta$) from \citet{2026AA...709A..56L} is shown in blue as an example of a low-gravity object in our sample, with our extracted SPHEREx spectrum shown as orange points. We also include the SPHEREx spectrum extracted with the SPIFF pipeline \citep{Gagne2026} in gray as comparison. The width of the lines indicates the uncertainty in the spectra. The most prominent molecular features found in L and T dwarfs for ${\rm CH_4}$, ${\rm CO_2}$, CO and ${\rm H_2O}$ are highlighted to indicate which features each JWST filter is capturing.} 
\label{fig:phot_filters}
\end{center}
\end{figure*}

For our study, we focus on the JWST photometric bandpasses that are especially sensitive to atmospheric properties of L and T dwarfs---including effective temperature, surface gravity, cloud properties, vertical mixing, and composition.
To identify these bands, we compare the most common molecular features present in the 1--5\,${\rm\mu m}$ range to the medium and wide filters from JWST/NIRCam, shown in Figure~\ref{fig:phot_filters}. We did not include the narrow filters, as the resolution of SPHEREx does not allow precise synthetic photometry for those filters.
Figure~\ref{fig:phot_filters} also shows the molecular features identified by \citet{Cushing2005} for ${\rm H_2O}$, ${\rm CH_4}$ and CO in the range 0.6--4.1\,${\rm \mu m}$, and by \citet{Yamamura2010} for CO and ${\rm CO_2}$ between 2.5--5.0\,${\rm \mu m}$, color-coded by molecule. 
We also highlight, as an example, the JWST NIRSpec Prism spectrum for 2MASS J22443167+2043433 (L7.5 $\beta$) from \citet{2026AA...709A..56L}, compared to our extraction from SPHEREx. 
The two spectra agree well, with a reduced chi-squared of $\chi^2_\nu=3.80$, and variations due to the object's variability \citep{Vos2018} and/or calibration issues. To confirm that the discrepancy does not originate in our extraction pipeline, we also compare our data with the SPHEREx spectrum generated with the SPIFF pipeline \citep{Gagne2026}, and find that both spectra agree, with $\chi^2_\nu=1.48$.

Here we focus on the filters that probe the fundamental molecular bands of key atmospheric species in the 3--5\,${\rm\mu m}$ range. In addition, we include one band (F410M) that is sensitive to the local continuum flux level, and two (F115W and F210M) others near $J$ and $K$ bands. To facilitate the comparison to imaged giant planets, we also added MKO $L'$ to our analysis. These filters are highlighted in Figure~\ref{fig:phot_filters} in bold text, and are summarized below together with their significance in this context:

\begin{itemize}
\item F115W overlaps with $J$ band, providing a reference to the traditional near-infrared filter scheme.
\item F210M samples the pseudocontinuum peak in  $K$-band, with minimal contamination from strong molecular absorption.
\item F335M probes the fundamental CH$_4$ $\nu_3$ band at 3.3 $\mu$m, one of the strongest methane absorption features in cool brown dwarfs and giant planets.
\item F410M samples a relatively transparent atmospheric window between the major CH$_4$, CO$_2$, and CO absorption bands, making it a useful proxy for the local continuum flux.
\item F430M probes the fundamental ${\rm CO_2}$ band near $4.3\,{\rm \mu m}$.
\item F444W spans both the CO$_2$ band near 4.3 $\mu$m and the blue wing of the fundamental CO band, making it sensitive to the combined effects of both species.
\item F460M probes the fundamental CO vibration-rotation band near 4.6 $\mu$m, a key tracer of carbon chemistry and vertical mixing.
\end{itemize}

For completeness, we also provide the synthetic photometry for all the JWST/NIRCam medium and wide filters in the 1--5\,${\rm \mu m}$ range, as well as the MKO $J$, $H$, $K$, $L'$ and $M'$ bands; all of these filters are marked in Figure~\ref{fig:phot_filters}. Synthetic photometry for all filters, along with spectral type, near-infrared gravity classification, and parallax for our samples of standards and low-gravity L and T dwarfs, can be found in Tables~\ref{table:phot}, \ref{table:phot2}, and \ref{table:phot3}.

\begin{deluxetable*}{cccccccccc}[ht!]
\tabletypesize{\scriptsize}
\setlength{\tabcolsep}{3pt}
\tablecaption{Synthetic photometry for low-gravity and standard L and T dwarfs \label{table:phot}}
\tablehead{\colhead{Name} &  \colhead{SpT} & \colhead{Gravity} &\colhead{Parallax}& \colhead{F115W}& \colhead{F140M}& \colhead{F150W}& \colhead{F162M}& \colhead{F182M}& \colhead{F200W}\\ & &  & (mas) & (mag) & (mag) & (mag) & (mag) & (mag) & (mag) }
\startdata 
2MASS J08095903+4434216 & L5.4$^{13}$ & INT-G$^{13}$ & ${42.40}\pm{3.60}^{27}$ & ${16.77}\pm{0.05}$ & ${16.33}\pm{0.05}$ & ${15.72}\pm{0.03}$ & ${15.18}\pm{0.03}$ & ${15.13}\pm{0.04}$ & ${14.70}\pm{0.03}$\\ 
2MASS J00191296-6226005 & L1.0$^{13}$ & VL-G$^{13}$ & ${20.60}\pm{1.67}^{30}$ & ${16.07}\pm{0.03}$ & ${15.67}\pm{0.03}$ & ${15.10}\pm{0.02}$ & ${14.57}\pm{0.02}$ & ${14.60}\pm{0.03}$ & ${14.30}\pm{0.02}$\\ 
DENIS J060852.8-275358 & L0.0$^{7}$ & VL-G$^{7}$ & ${22.62}\pm{0.16}^{30}$ & ${13.93}\pm{0.02}$ & ${13.66}\pm{0.02}$ & ${13.29}\pm{0.02}$ & ${12.84}\pm{0.02}$ & ${12.87}\pm{0.02}$ & ${12.66}\pm{0.02}$\\ 
2MASS J01205114-5200349 & L1.4$^{13}$ & VL-G$^{13}$ & ${24.26}\pm{0.94}^{30}$ & ${16.05}\pm{0.03}$ & ${15.75}\pm{0.03}$ & ${15.14}\pm{0.02}$ & ${14.57}\pm{0.02}$ & ${14.59}\pm{0.02}$ & ${14.29}\pm{0.02}$\\ 
2MASS J00182834-6703130 & L0.4$^{13}$ & VL-G$^{13}$ & ${22.00}\pm{0.94}^{30}$ & ${15.75}\pm{0.03}$ & ${15.35}\pm{0.03}$ & ${14.89}\pm{0.02}$ & ${14.37}\pm{0.02}$ & ${14.34}\pm{0.02}$ & ${14.08}\pm{0.02}$\\ 
2MASSI J0443376+000205 & L0.0$^{7}$ & VL-G$^{7}$ & ${47.62}\pm{0.14}^{30}$ & ${12.75}\pm{0.02}$ & ${12.46}\pm{0.02}$ & ${12.11}\pm{0.02}$ & ${11.70}\pm{0.02}$ & ${11.71}\pm{0.02}$ & ${11.53}\pm{0.02}$\\ 
2MASS J05445741+3705039 & L1.9$^{32}$ & INT-G$^{32}$ & ${34.38}\pm{0.33}^{30}$ & ${14.27}\pm{0.02}$ & ${13.96}\pm{0.02}$ & ${13.46}\pm{0.02}$ & ${12.96}\pm{0.02}$ & ${13.01}\pm{0.02}$ & ${12.73}\pm{0.02}$\\ 
SDSS J154705.54-162631.6 & M9.4$^{14}$ & INT-G$^{14}$ & ${26.73}\pm{0.34}^{30}$ & ${14.14}\pm{0.02}$ & ${13.86}\pm{0.02}$ & ${13.54}\pm{0.02}$ & ${13.19}\pm{0.02}$ & ${13.19}\pm{0.02}$ & ${13.02}\pm{0.02}$\\ 
2MASS J02410564-5511466 & L1.6$^{13}$ & VL-G$^{13}$ & ${23.86}\pm{0.80}^{30}$ & ${15.68}\pm{0.03}$ & ${15.36}\pm{0.03}$ & ${14.82}\pm{0.02}$ & ${14.29}\pm{0.02}$ & ${14.35}\pm{0.02}$ & ${14.09}\pm{0.02}$\\ 
DENIS J022519.4-583729 & M9.4$^{17}$ & INT-G$^{17}$ & ${24.25}\pm{0.17}^{30}$ & ${13.97}\pm{0.02}$ & ${13.73}\pm{0.02}$ & ${13.39}\pm{0.02}$ & ${13.00}\pm{0.02}$ & ${13.04}\pm{0.02}$ & ${12.85}\pm{0.02}$\\ 
\enddata 
\tablecomments{First ten objects of our sample. The full table can be found online \url{https://zenodo.org/records/21856313} \citep{kiman_2026_21856313}. All magnitudes are reported in the Vega magnitude system.}
\tablerefs{[1] \citet{2004AJ....127.3553K}; [2] \citet{2006ApJ...637.1067B}; [3] \citet{2006ApJ...645..676L}; [4] \citet{2010ApJS..190..100K}; [5] \citet{2012ApJS..201...19D}; [6] \citet{2013AJ....146..161M}; [7] \citet{2013ApJ...772...79A}; [8] \citet{2013ApJ...777L..20L}; [9] \citet{2014ApJ...785L..14G}; [10] \citet{2014ApJ...794..143B}; [11] \citet{2015AJ....150..182K}; [12] \citet{2015ApJ...814..118B}; [13] \citet{2015ApJS..219...33G}; [14] \citet{2016ApJ...821..120A}; [15] \citet{2016ApJ...822L...1S}; [16] \citet{2016ApJ...833...96L}; [17] \citet{2016ApJS..225...10F}; [18] \citet{2016PhDT.......189A}; [19] \citet{2017AJ....153..196S}; [20] \citet{2017AJ....154...69S}; [21] \citet{2017AJ....154..112K}; [22] \citet{2017AJ....154..147D}; [23] \citet{2017ApJ...837...95B}; [24] \citet{2017ApJS..228...18G}; [25] \citet{2020AA...633A.152C}; [26] \citet{2020AJ....159..257B}; [27] \citet{2021ApJS..253....7K}; [28] \citet{2022ApJ...924...68V}; [29] \citet{2022ApJ...934..178A}; [30] \citet{2023AA...674A...1G}; [31] \citet{2023ApJ...959...63S}; [32] \citet{2024ApJ...961..121H}; [33] \citet{2024ApJ...967..115B}; [34] \citet{2026AA...709A..56L}; 
}\end{deluxetable*}

\begin{deluxetable*}{ccccccccc}[ht!]
\tabletypesize{\scriptsize}
\setlength{\tabcolsep}{3pt}
\tablecaption{Synthetic photometry for low-gravity and standard L and T dwarfs (continued) \label{table:phot2}}
\tablehead{\colhead{Name} & \colhead{F210M}& \colhead{F250M}& \colhead{F277W}& \colhead{F300M}& \colhead{F335M}& \colhead{F356W}& \colhead{F360M}& \colhead{F410M}\\  & (mag) & (mag) & (mag) & (mag) & (mag) & (mag) & (mag) & (mag) }
\startdata 
2MASS J08095903+4434216 & ${14.37}\pm{0.03}$ & ${14.22}\pm{0.04}$ & ${14.01}\pm{0.03}$ & ${13.72}\pm{0.03}$ & ${13.20}\pm{0.03}$ & ${13.03}\pm{0.02}$ & ${12.93}\pm{0.03}$ & ${12.81}\pm{0.03}$\\ 
2MASS J00191296-6226005 & ${14.06}\pm{0.02}$ & ${13.97}\pm{0.03}$ & ${13.87}\pm{0.02}$ & ${13.69}\pm{0.03}$ & ${13.22}\pm{0.03}$ & ${13.05}\pm{0.02}$ & ${12.96}\pm{0.03}$ & ${12.89}\pm{0.03}$\\ 
DENIS J060852.8-275358 & ${12.48}\pm{0.02}$ & ${12.47}\pm{0.02}$ & ${12.43}\pm{0.02}$ & ${12.32}\pm{0.02}$ & ${11.89}\pm{0.02}$ & ${11.72}\pm{0.02}$ & ${11.63}\pm{0.02}$ & ${11.52}\pm{0.02}$\\ 
2MASS J01205114-5200349 & ${14.06}\pm{0.02}$ & ${13.98}\pm{0.03}$ & ${13.83}\pm{0.02}$ & ${13.62}\pm{0.02}$ & ${13.15}\pm{0.02}$ & ${12.98}\pm{0.02}$ & ${12.89}\pm{0.02}$ & ${12.80}\pm{0.03}$\\ 
2MASS J00182834-6703130 & ${13.87}\pm{0.02}$ & ${13.74}\pm{0.03}$ & ${13.60}\pm{0.02}$ & ${13.43}\pm{0.03}$ & ${13.06}\pm{0.02}$ & ${12.93}\pm{0.02}$ & ${12.84}\pm{0.02}$ & ${12.76}\pm{0.03}$\\ 
2MASSI J0443376+000205 & ${11.34}\pm{0.02}$ & ${11.36}\pm{0.02}$ & ${11.30}\pm{0.02}$ & ${11.17}\pm{0.02}$ & ${10.74}\pm{0.02}$ & ${10.58}\pm{0.02}$ & ${10.49}\pm{0.02}$ & ${10.38}\pm{0.02}$\\ 
2MASS J05445741+3705039 & ${12.50}\pm{0.02}$ & ${12.52}\pm{0.03}$ & ${12.46}\pm{0.02}$ & ${12.28}\pm{0.02}$ & ${11.81}\pm{0.02}$ & ${11.62}\pm{0.02}$ & ${11.54}\pm{0.02}$ & ${11.46}\pm{0.02}$\\ 
SDSS J154705.54-162631.6 & ${12.86}\pm{0.02}$ & ${12.94}\pm{0.02}$ & ${12.86}\pm{0.02}$ & ${12.73}\pm{0.02}$ & ${12.35}\pm{0.02}$ & ${12.21}\pm{0.02}$ & ${12.15}\pm{0.02}$ & ${12.03}\pm{0.02}$\\ 
2MASS J02410564-5511466 & ${13.87}\pm{0.02}$ & ${13.87}\pm{0.03}$ & ${13.71}\pm{0.02}$ & ${13.52}\pm{0.02}$ & ${13.05}\pm{0.02}$ & ${12.88}\pm{0.02}$ & ${12.78}\pm{0.02}$ & ${12.73}\pm{0.02}$\\ 
DENIS J022519.4-583729 & ${12.68}\pm{0.02}$ & ${12.75}\pm{0.02}$ & ${12.71}\pm{0.02}$ & ${12.59}\pm{0.02}$ & ${12.17}\pm{0.02}$ & ${12.00}\pm{0.02}$ & ${11.91}\pm{0.02}$ & ${11.81}\pm{0.02}$\\ 
\enddata 
\tablecomments{See notes from Table~\ref{table:phot}.}\end{deluxetable*}

\begin{deluxetable*}{ccccccccc}[ht!]
\tabletypesize{\scriptsize}
\setlength{\tabcolsep}{3pt}
\tablecaption{Synthetic photometry for low-gravity and standard L and T dwarfs (continued)\label{table:phot3}}
\tablehead{\colhead{Name} & \colhead{F430M}& \colhead{F444W}& \colhead{F460M}& \colhead{$J_{\rm MKO}$}& \colhead{$H_{\rm MKO}$}& \colhead{$K_{\rm MKO}$}& \colhead{$L'_{\rm MKO}$}& \colhead{$M'_{\rm MKO}$}\\  & (mag) & (mag) & (mag) & (mag) & (mag) & (mag) & (mag) & (mag) }
\startdata 
2MASS J08095903+4434216 & ${12.94}\pm{0.04}$ & ${12.92}\pm{0.03}$ & ${13.10}\pm{0.06}$ & ${16.38}\pm{0.05}$ & ${15.30}\pm{0.03}$ & ${14.22}\pm{0.03}$ & ${12.88}\pm{0.02}$ & ${13.06}\pm{0.06}$\\ 
2MASS J00191296-6226005 & ${12.94}\pm{0.03}$ & ${12.95}\pm{0.03}$ & ${13.03}\pm{0.04}$ & ${15.61}\pm{0.03}$ & ${14.68}\pm{0.02}$ & ${13.88}\pm{0.02}$ & ${12.94}\pm{0.02}$ & ${12.98}\pm{0.04}$\\ 
DENIS J060852.8-275358 & ${11.59}\pm{0.03}$ & ${11.63}\pm{0.02}$ & ${11.75}\pm{0.03}$ & ${13.57}\pm{0.02}$ & ${12.93}\pm{0.02}$ & ${12.29}\pm{0.02}$ & ${11.60}\pm{0.02}$ & ${11.72}\pm{0.03}$\\ 
2MASS J01205114-5200349 & ${12.90}\pm{0.03}$ & ${12.89}\pm{0.02}$ & ${12.96}\pm{0.04}$ & ${15.61}\pm{0.02}$ & ${14.68}\pm{0.02}$ & ${13.86}\pm{0.02}$ & ${12.85}\pm{0.02}$ & ${12.93}\pm{0.04}$\\ 
2MASS J00182834-6703130 & ${12.81}\pm{0.04}$ & ${12.80}\pm{0.02}$ & ${12.80}\pm{0.03}$ & ${15.32}\pm{0.02}$ & ${14.47}\pm{0.02}$ & ${13.67}\pm{0.02}$ & ${12.83}\pm{0.02}$ & ${12.81}\pm{0.04}$\\ 
2MASSI J0443376+000205 & ${10.46}\pm{0.02}$ & ${10.48}\pm{0.02}$ & ${10.56}\pm{0.02}$ & ${12.41}\pm{0.02}$ & ${11.77}\pm{0.02}$ & ${11.17}\pm{0.02}$ & ${10.45}\pm{0.02}$ & ${10.54}\pm{0.02}$\\ 
2MASS J05445741+3705039 & ${11.55}\pm{0.03}$ & ${11.58}\pm{0.02}$ & ${11.67}\pm{0.03}$ & ${13.88}\pm{0.02}$ & ${13.06}\pm{0.02}$ & ${12.32}\pm{0.02}$ & ${11.51}\pm{0.02}$ & ${11.66}\pm{0.03}$\\ 
SDSS J154705.54-162631.6 & ${12.09}\pm{0.03}$ & ${12.14}\pm{0.02}$ & ${12.23}\pm{0.03}$ & ${13.86}\pm{0.02}$ & ${13.25}\pm{0.02}$ & ${12.72}\pm{0.02}$ & ${12.10}\pm{0.02}$ & ${12.22}\pm{0.03}$\\ 
2MASS J02410564-5511466 & ${12.86}\pm{0.03}$ & ${12.83}\pm{0.03}$ & ${12.86}\pm{0.04}$ & ${15.23}\pm{0.02}$ & ${14.40}\pm{0.02}$ & ${13.70}\pm{0.02}$ & ${12.76}\pm{0.02}$ & ${12.84}\pm{0.04}$\\ 
DENIS J022519.4-583729 & ${11.88}\pm{0.02}$ & ${11.94}\pm{0.02}$ & ${12.08}\pm{0.03}$ & ${13.67}\pm{0.02}$ & ${13.08}\pm{0.02}$ & ${12.52}\pm{0.02}$ & ${11.87}\pm{0.02}$ & ${12.06}\pm{0.03}$\\ 
\enddata 
\tablecomments{See notes from Table~\ref{table:phot}.}\end{deluxetable*}

\subsection{Color and absolute magnitude versus spectral type}
\label{subsec:colorabsvsspt}

\begin{figure*}[ht!]
\begin{center}
\includegraphics[width=\linewidth]{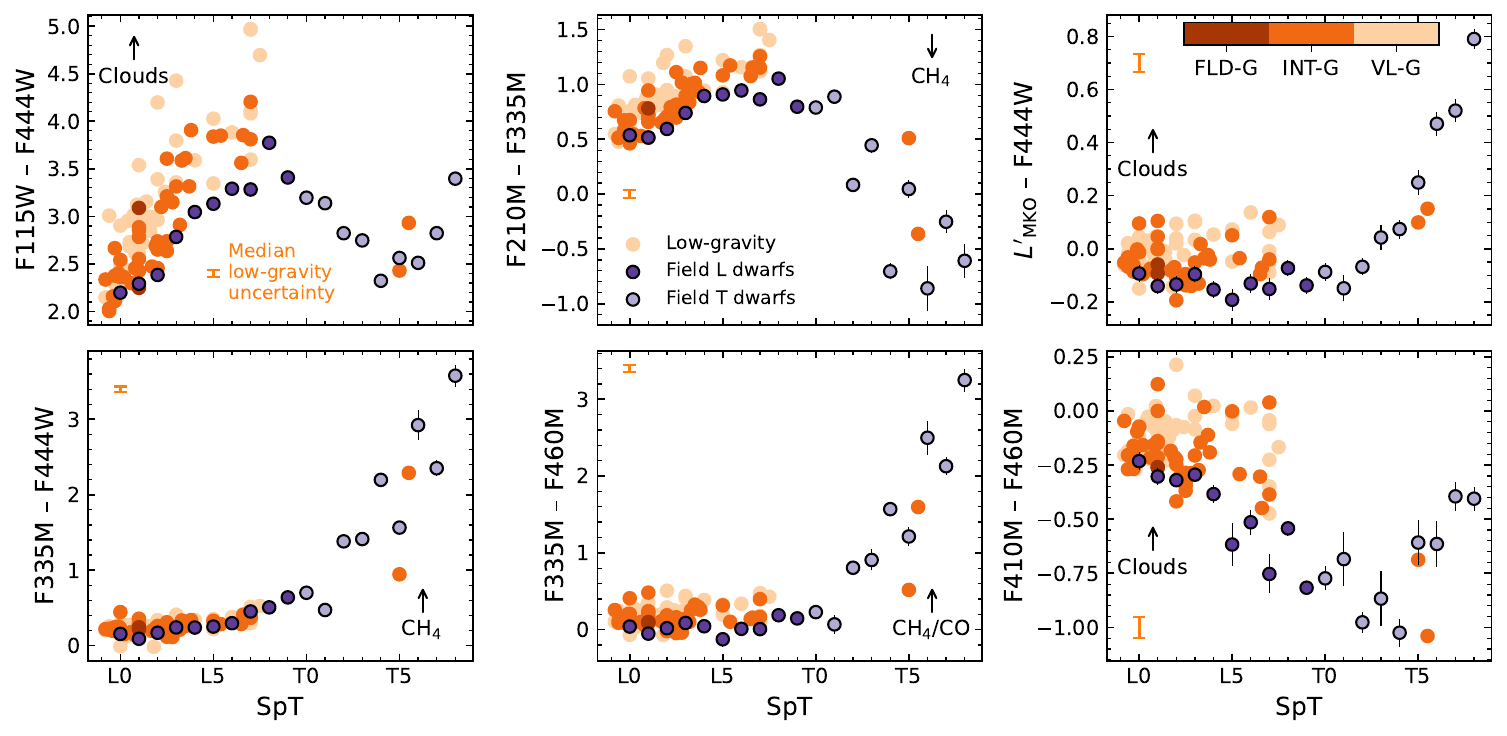}
\caption{Synthetic colors using JWST/NIRCam filters, calculated from SPHEREx spectra, for field dwarfs (L dwarfs in dark purple and T dwarfs in light purple) and low-gravity L and T dwarfs (color-coded by gravity classification). Colors are shown as a function of infrared spectral type. The median uncertainty for low-gravity objects is shown in orange. We also include, in black, the physical interpretation of each color, with an arrow indicating the direction of increase for that parameter.} 
\label{fig:JWST_phot_color_spt}
\end{center}
\end{figure*}

To investigate the diversity of low-gravity objects, we examine a set of JWST/NIRCam colors selected to probe strong spectral features and analyze their behavior as a function of spectral type. Figure~\ref{fig:JWST_phot_color_spt} presents six representative colors spanning the 1--5\,$\mu$m wavelength range. Each panel includes our sample of low-gravity L and T dwarfs, color-coded by their near-infrared gravity classification, and field L and T dwarf in dark and light purple, respetively.

Across these diagrams, low-gravity L dwarfs exhibit systematically redder 3--5\,$\mu$m colors than their field counterparts. This behavior is broadly consistent with differences in cloud properties associated with low surface gravity and extends to thermal-infrared wavelengths the cloud-related color trends previously identified in the near infrared. The persistence of these offsets over a broad wavelength range demonstrates that the atmospheric effects of low surface gravity remain significant well beyond the traditional near-infrared diagnostics. We discuss each panel in turn below.

\subsubsection{F115W--F444W}

The (F115W--F444W) color provides one of the strongest empirical diagnostics of low gravity in our sample. It measures the slope of the spectral energy distribution from the $J$ band to the mid-infrared and clearly separates field and low-gravity L dwarfs, as expected from the spectral differences shown in Figures~\ref{fig:spectra_comparison0} and \ref{fig:spectra_comparison1}. Field objects become progressively redder from early- to late-L spectral types as decreasing effective temperature leads to thicker silicate and iron clouds. These clouds increase the opacity in the near-infrared windows ($z$, $J$, and $H$), suppressing the emergent flux from deep, hot atmospheric layers. The absorbed flux is redistributed to longer wavelengths where the atmosphere is more transparent, producing increasingly red near-to-mid-infrared colors. Across the L/T transition, the color becomes abruptly bluer as condensate clouds settle below the photosphere and no longer suppress the near-infrared flux windows \citep{Ackerman2001,Burgasser2002,Kirkpatrick2005,Burrows2006}.

Low-gravity objects follow the same overall evolution with spectral type but are systematically redder than field dwarfs. This offset is consistent with clouds residing higher in the photosphere and having larger optical depths, leading to greater suppression of the near-infrared flux and enhanced emission at longer wavelengths \citep{2013ApJ...772...79A,2016ApJS..225...10F,2016ApJ...833...96L,2021ApJS..253....7K}. A clear dependence on gravity classification is also present, with VL-G objects generally redder than INT-G and FLD-G objects of the same spectral type. Despite this trend, there remains substantial intrinsic scatter of approximately 1 mag within each gravity class, indicating that surface gravity alone does not determine the emergent colors of young brown dwarfs. This diversity may reflect variations in cloud properties, patchy or evolving cloud structures \citep[e.g.,][]{Apai2013}, viewing geometry \citep{Vos2020,Suarez2023}, metallicity \citep{Balmer2026}, or other atmospheric parameters.

\subsubsection{F210M--F335M}

The (F210M--F335M) color provides a sensitive probe of the onset and strengthening of the fundamental CH$_4$ absorption band at 3.3 $\mu$m relative to the nearby $K$-band continuum. Field L dwarfs become progressively redder from L0 to L7 as increasingly thick condensate clouds suppress the $K$-band flux while having a smaller effect at longer wavelengths. Consequently, the reddening is similar to that seen in (F115W--F444W), although with a smaller amplitude because cloud opacity has less influence at 2 $\mu$m than at shorter wavelengths. Across the L/T transition, the color becomes rapidly bluer as condensate clouds settle below the photosphere, increasing the F210M flux, while the emergence of strong methane absorption suppresses the F335M flux.
Low-gravity objects broadly follow the same evolution with spectral type but display a substantial spread of approximately 0.5 mag at fixed subtype. In contrast to (F115W--F444W), however, the separation between field and low-gravity objects is less pronounced, indicating that this color is governed by the competing effects of cloud opacity and methane absorption. The remaining scatter likely reflects differences in cloud properties, atmospheric chemistry, and other secondary atmospheric parameters.

\subsubsection{$L$'--F444W}

($L$'--F444W) probes the strength of the fundamental CO absorption band at 4.6$\,\mu$m relative to the adjacent continuum sampled by F444W. Low-gravity objects exhibit a modest offset from the field sequence, although the interpretation is not straightforward. While this behavior could be attributed to weaker apparent CO absorption, atmospheric models do not predict substantially lower CO abundances in low-gravity atmospheres \citep{mukherjee_2025_15150881}. Instead, the observed offset may reflect the effects of thicker condensate clouds, whose additional continuum opacity shifts the photosphere to higher, cooler atmospheric layers and reduces the apparent contrast of molecular absorption bands.

\subsubsection{F335M--F444W}

The (F335M--F444W) color directly traces the strength of the fundamental CH$_4$ absorption band at 3.3 $\mu$m. The F335M filter samples the center of the methane feature, while F444W lies largely outside the band and serves as a comparison to the longer-wavelength flux \citep{Cushing2005,mukherjee_2025_15150874}. As methane absorption strengthens with decreasing effective temperature, the F335M flux is increasingly suppressed, producing progressively redder colors from early- to late-T dwarfs. Similar behavior is evident in the (F335M--F410M) and (F335M--F430M) colors.
In contrast to colors spanning the near- and thermal-infrared, low-gravity objects do not exhibit a significant systematic offset from the field sequence. Because all of these filters sample wavelengths between 3 and 5\,$\mu$m, where the differential effects of condensate clouds are comparatively modest, the colors are governed primarily by the strength of methane absorption. The remaining scatter likely reflects variations in atmospheric chemistry, vertical mixing, and other secondary atmospheric properties that influence the CH$_4$ abundance.

\subsubsection{F335M--F460M}

The (F335M--F460M) color is closely related to (F335M--F444W), but provides slightly better separation between low-gravity and field objects. This color compares the fundamental CH$_4$ absorption band at 3.3~$\mu$m with the fundamental CO band at 4.6~$\mu$m, making it sensitive to the balance between these two carbon-bearing species. Redder colors indicate relatively stronger CH$_4$ absorption or weaker CO absorption, while bluer colors indicate the opposite.

\subsubsection{F410M--F460M}

(F410M--F460M) compares the local continuum at $4.1\,{\rm \mu m}$ (with a slight overlap with the CO$_2$ absorption feature at $4.3\,{\rm \mu m}$) with the CO fundamental absorption at 4.67$\,\mu$m, making it sensitive to the relative strengths of these carbon-bearing molecules. As with ($L$'--F444W), low-gravity objects are modestly offset from the field sequence. This behavior is consistent with enhanced cloud opacity muting molecular absorption features by shifting the photosphere to higher, cooler atmospheric layers, rather than indicating substantially different CO or CO$_2$ abundances \citep{mukherjee_2025_15150881}. Similar trends are present in the (F430M--F460M), (F410M--F444W), (F430M--F444W), and (F444W--F460M) colors.

\subsubsection{Absolute Magnitudes as a Function of Spectral Type}

\begin{figure*}[ht!]
\begin{center}
\includegraphics[width=\linewidth]{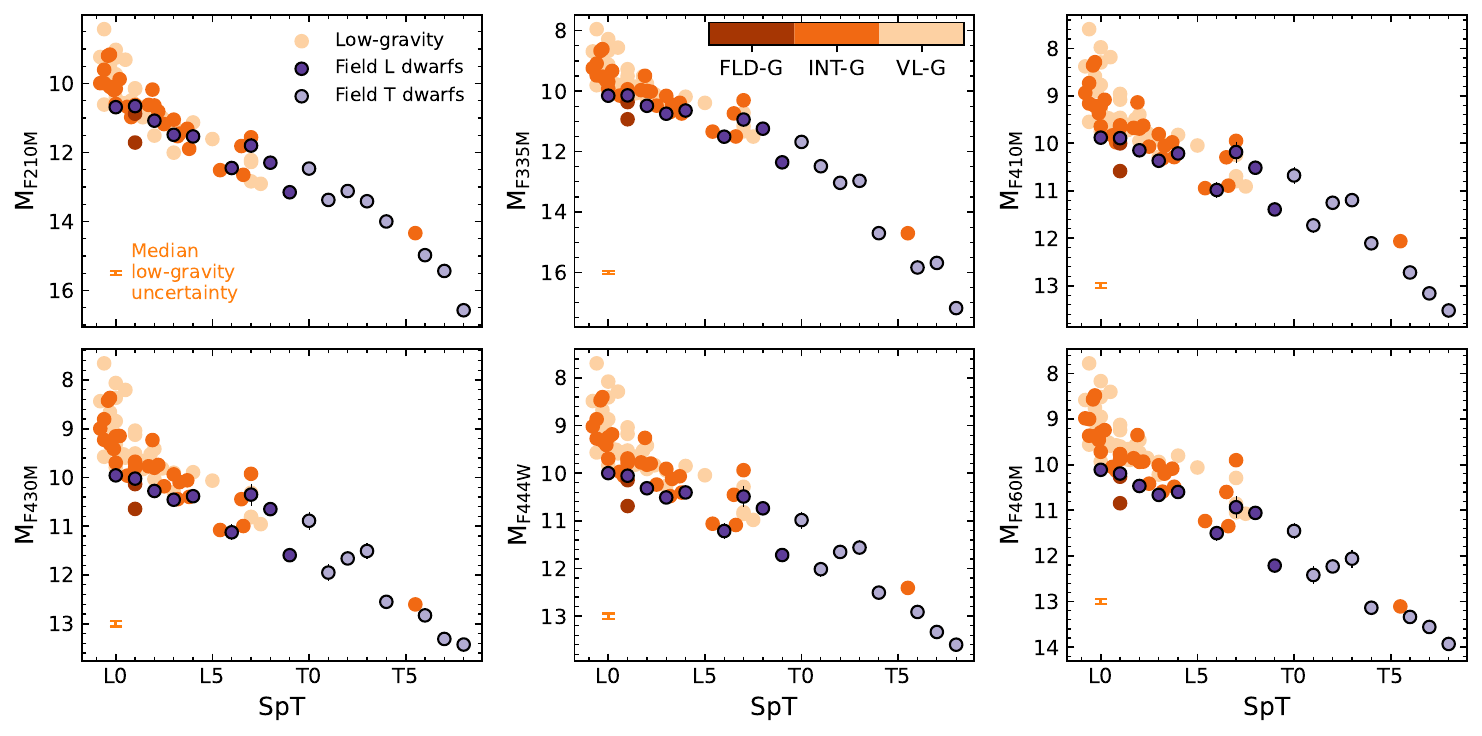}
\caption{Synthetic absolute magnitudes using a selection of JWST/NIRCam filters, calculated from SPHEREx spectra, for field dwarfs (L dwarfs in dark purple and T dwarfs in light purple) and low-gravity L and T dwarfs (color-coded by gravity classification). Parallaxes for each source are obtained from the UltracoolSheet. Absolute magnitudes are shown as a function of infrared spectral type. The median uncertainty for low-gravity objects is shown in orange.} 
\label{fig:JWST_phot_absmag_spt}
\end{center}
\end{figure*}

Parallaxes are available for 68 of the 94 objects in our sample from the UltracoolSheet. We use these parallaxes to calculate absolute magnitudes in the F210M, F335M, F410M, F430M, F444W, and F460M filters and examine their evolution with near-infrared spectral type. Figure~\ref{fig:JWST_phot_absmag_spt} shows the resulting sequences. We adopt the same plotting conventions as in Figure~\ref{fig:JWST_phot_color_spt}, including the low-gravity objects and the field standards described in Section~\ref{subsec:standards}.

In all six filters, the field sequence becomes progressively fainter toward later spectral types, broadly tracing the decline in bolometric luminosity with decreasing effective temperature. Low-gravity objects are systematically overluminous relative to the field sequence at a given spectral type, although substantial scatter is present. Similar offsets have previously been identified from the near-infrared through approximately $4.5\,\mu$m using primarily 2MASS and WISE photometry \citep{Filippazzo2015,2016ApJS..225...10F,2021ApJS..253....7K}. Our results show that this behavior persists across the full set of JWST filters considered here.

The enhanced luminosities of low-gravity objects likely arise in part from their youth and larger radii. At a fixed effective temperature, an inflated radius produces a higher bolometric luminosity and therefore brighter absolute magnitudes. However, the interpretation at fixed spectral type is more complicated because the relation between spectral type and effective temperature depends on surface gravity. Differences in cloud opacity and atmospheric structure can also redistribute flux across the 2--5~$\mu$m region, contributing to both the mean offset and the observed scatter about the field sequence.

\subsection{How similar are low-gravity L dwarfs and young giant planets in the near and thermal infrared?}

Low-gravity L and T dwarfs share similar effective temperatures, masses and ages with directly imaged young exoplanets. Here we use our samples of low-gravity and field objects to compare their color-color and color-magnitude locus to imaged exoplanets which have either a 1--5\,${\rm \mu m}$ JWST spectrum---for which we synthesize relevant photometry and colors---or actual photometric measurements with subsets of NIRCam filters. By using our measured photometry, we obtain an empirical low-gravity sequence, independent of models. We include in our analysis the JWST/NIRCam photometry of HR~8799~bcde \citep{Marois2008,Marois2010,Balmer2025}, 29~Cyg~b \citep[HIP~99770~b,][]{Currie2023,Balmer2026}, HIP~65426~b \citep{2017A&A...605L...9C,2023ApJ...951L..20C} and $\beta$~Pic~b and d \citep{Lagrange2009,Lagrange2010,Stolker2019,Kammerer2024,Sutlieff2026,Gibbs2026}, as well as synthesized photometry of COCONUTS-2~b \citep{Kirkpatrick2011,2021ApJ...916L..11Z,Kiman2026}, TWA~27~b \citep{Chauvin2004,Manjavacas2024} and VHS~1256--1257~b \citep{Gauza2015,Miles2023} based on their JWST/NIRSpec spectra. For comparison, we also include the field L and T dwarfs from this work, together with field T and Y dwarfs from \citet{Beiler2024} to extend the sequence to cooler spectral types.

\subsubsection{Color-Color Diagrams}

\begin{figure*}[ht!]
\begin{center}
\includegraphics[width=\linewidth]{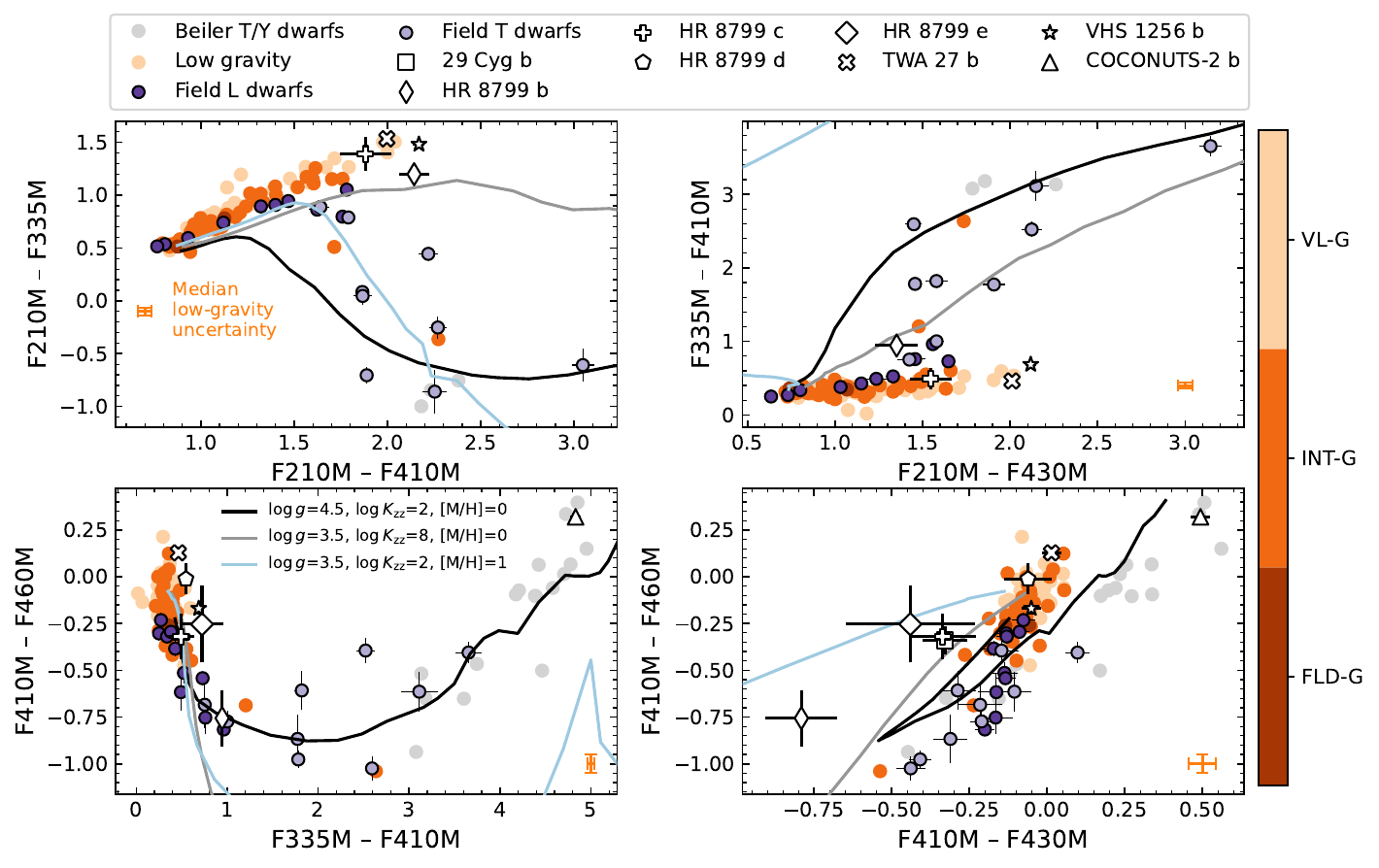}
\caption{Color-color plots for the synthetic photometry for JWST/NIRCam filters, for the low-gravity and standard samples obtained from the SPHEREx spectra. We use the same color-coding convention as for Figures~\ref{fig:JWST_phot_color_spt} and \ref{fig:JWST_phot_absmag_spt}. In orange, we show the median error-bars for the low-gravity objects in each subplot. Field L and T dwarfs from this work are shown in dark and light purple, respectively, and from \citet{Beiler2024} in gray. We include only the objects from \citet{Beiler2024} that fit within the ranges of our data. In addition, we included the photometry of the giant planets HR 8799 bcde \citep{Balmer2025}, 29 Cyg b \citep{Balmer2026}, COCONUTS-2 b \citep{Kiman2026}, TWA 27 b \citep{Manjavacas2024} and VHS~1256--1257~b \citep{Miles2023} as comparison for the low-gravity objects. We also show the Sonora Elf Owl models, for the color-color plots for three cases: 1) $\log g=4.5$, $\log K_{\rm zz}=2$ and $[{\rm M/H}]=0$; 2) $\log g=3.5$, $\log K_{\rm zz}=8$ and $[{\rm M/H}]=0$; 3) $\log g=3.5$, $\log K_{\rm zz}=2$ and $[{\rm M/H}]=1$. These three cases cover low and high gravity and vertical mixing, and solar and high-metallicity.} 
\label{fig:JWST_phot}
\end{center}
\end{figure*}

We select four color-color diagrams, shown in Figure~\ref{fig:JWST_phot}, that are especially informative to highlight the photometric properties of imaged exoplanets. We include the sample of low-gravity and standard L and T dwarfs with the same conventions as in Figures~\ref{fig:JWST_phot_color_spt} and \ref{fig:JWST_phot_absmag_spt}, and the Sonora Elf Owl models \citep{mukherjee_2025_15150881,mukherjee_2025_15150874,2016JQSRT.177...15K,2020ApJS..247...55H,2021ApJ...923..269K,2021ApJ...920...85M,2023ApJ...942...71M} for three cases: 1) Field gravity, low vertical mixing and solar metallicity ($\log g=4.5$, $\log K_{\rm zz}=2$ and $[{\rm M/H}]=0$); 2) Low-gravity, high vertical mixing, solar metallicity ($\log g=3.5$, $\log K_{\rm zz}=8$ and $[{\rm M/H}]=0$); 3) Low-gravity, low vertical mixing, high metallicity ($\log g=3.5$, $\log K_{\rm zz}=2$ and $[{\rm M/H}]=1$). Below we describe each of the four panels from Figure~\ref{fig:JWST_phot}.

The (F210M--F335M) versus (F210M--F410M) diagram (top-left) provides one of the clearest empirical separations between field and low-gravity objects. The (F210M--F335M) color traces the onset of the fundamental CH$_4$ absorption band at $3.3\,\mu$m, while (F210M--F410M) measures the overall spectral slope between two relatively transparent atmospheric windows. The low-gravity sequence extends systematically toward redder colors in both axes, with the VL-G objects defining the reddest locus. Most of the warm directly imaged planets considered here, including TWA~27~b, VHS~1256--1257~b, and HR~8799~c, follow this extension of the low-gravity sequence. HR~8799~b deviates slightly towards cooler temperatures. The cloud-free models reproduce the general direction of the observed trends with decreasing gravity, increasing metallicity, and enhanced vertical mixing, but fail to reproduce the full extent of the observed reddening, emphasizing the importance of condensate clouds in shaping the 2--5~$\mu$m spectral energy distribution.

A similar separation is evident in the (F335M--F410M) versus (F210M--F430M) diagram (top-right), which combines diagnostics of the CH$_4$ and CO$_2$ absorption bands. TWA~27~b, VHS~1256--1257~b, and HR~8799~c overlap or extend the empirical low-gravity sequence. HR~8799~b is the principal exception, lying closer to the field T-dwarf sequence and the high-metallicity models. This displacement is expected because HR~8799~b is approximately 200~K cooler than HR~8799~cde \citep{Nasedkin2024}, resulting in substantially stronger 3.3~$\mu$m methane absorption \citep{Xuan2026}.

The lower panels probe the thermal-infrared carbon chemistry. The (F410M--F460M) versus (F335M--F410M) diagram (bottom-left) shows considerably less separation between field and low-gravity objects than the previous color combinations, indicating that these colors are less sensitive to gravity than to the relative strengths of the CO and CH$_4$ absorption bands. Nevertheless, most directly imaged planets (TWA~27~b, VHS~1256--1257~b, and HR~8799~cde) remain consistent with the low-gravity locus. HR~8799~b again follows the cooler T-dwarf sequence because of its stronger methane absorption, while COCONUTS-2~b agrees well with the late-T field sequence, consistent with its classification as an older ($\sim$400 Myr) T9 companion \citep{Kiman2026}.

Finally, the (F410M--F460M) versus (F410M--F430M) diagram (bottom-right) compares colors sensitive to the relative strengths of the CO and CO$_2$ absorption bands. This color combination has been proposed as a diagnostic of enhanced atmospheric metallicity and vigorous vertical mixing \citep{Balmer2025,Balmer2026}. The low-gravity sequence remains offset from the field dwarfs, although the observational uncertainties are larger than in the previous panels. Several directly imaged planets, including HR~8799~bce and 29~Cyg~b, lie toward the region occupied by models with enhanced metallicity or stronger vertical mixing, whereas TWA~27~b, VHS~1256--1257~b, and HR~8799~d agree with low-gravity objects, and COCONUTS-2~b remains consistent with the late-T dwarf sequence. Overall, the color--color diagrams demonstrate that young giant planets generally follow the empirical sequence defined by free-floating low-gravity brown dwarfs while highlighting the importance of clouds, chemistry, and atmospheric mixing in shaping their 2--5~$\mu$m colors.

\subsubsection{Color-Magnitude Diagrams}

\begin{figure*}[ht!]
\begin{center}
\includegraphics[width=\linewidth]{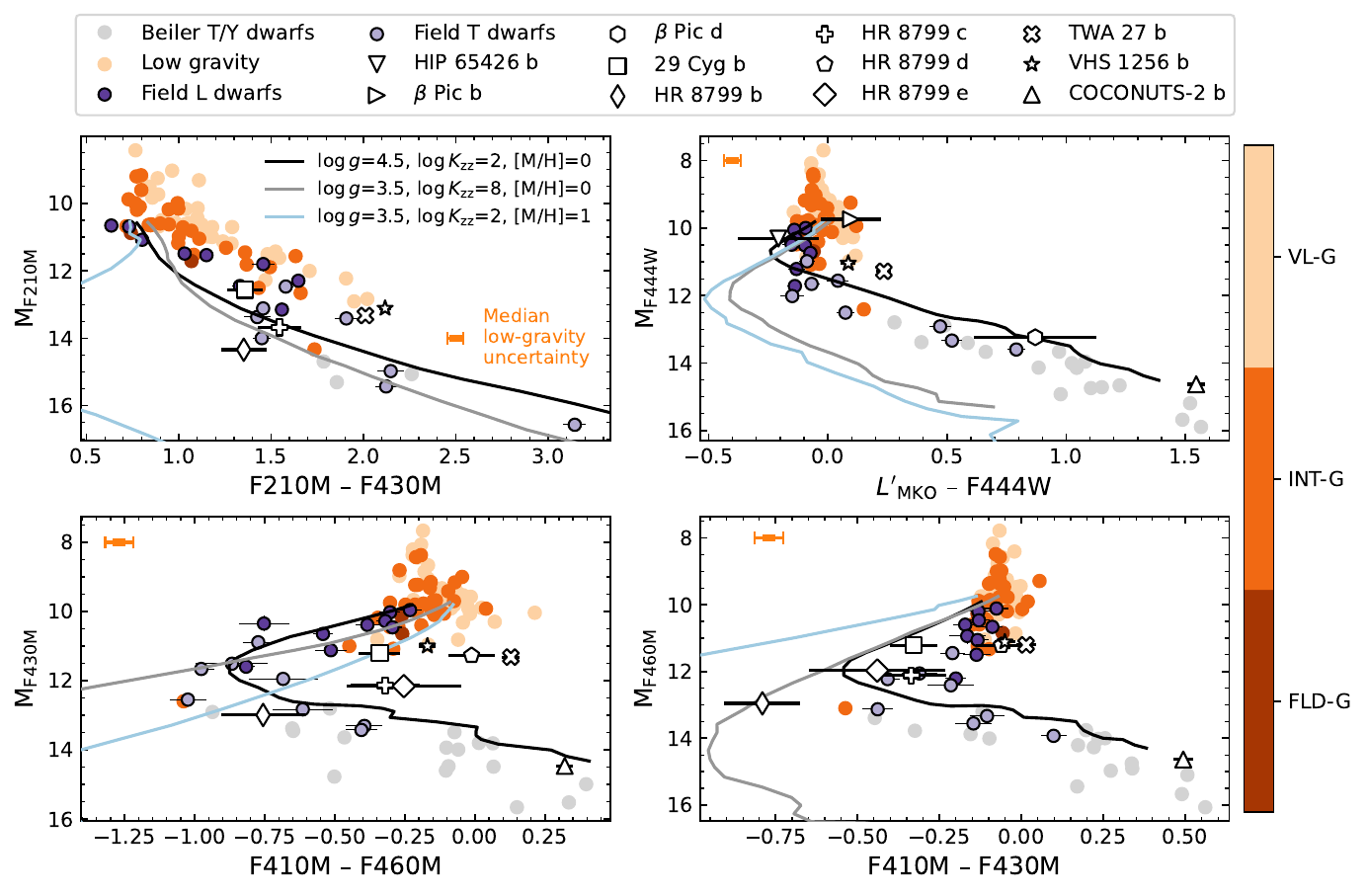}
\caption{Color-magnitude plots for the synthetic photometry for JWST/NIRCam filters, for the low-gravity and standard samples obtained from the SPHEREx spectra. In orange, we show the median error-bars for the low-gravity objects in each subplot. We included field L and T dwarfs from this work in dark and light purple, respectively, and from \citet{Beiler2024} in gray. We included only the objects from \citet{Beiler2024} that fit within the ranges of our data. We use the same color-coding convention as for Figures~\ref{fig:JWST_phot_color_spt} and \ref{fig:JWST_phot_absmag_spt}. In addition, we included the photometry of the giant planets HR~8799~bcde \citep{Balmer2025}, 29~Cyg~b \citep{Balmer2026}, HIP~65426~b \citep{2023ApJ...951L..20C}, $\beta$~Pic~bd \citep{Kammerer2024,Sutlieff2026,Gibbs2026}, COCONUTS-2 b \citep{Kiman2026}, TWA 27 b \citep{Manjavacas2024} and VHS~1256--1257~b \citep{Miles2023} as comparison for the low-gravity objects. We also show the Sonora Elf Owl models for the color-magnitude plots for three cases: 1) $\log g=4.5$, $\log K_{\rm zz}=2$ and $[{\rm M/H}]=0$; 2) $\log g=3.5$, $\log K_{\rm zz}=8$ and $[{\rm M/H}]=0$; 3) $\log g=3.5$, $\log K_{\rm zz}=2$ and $[{\rm M/H}]=1$. These three cases cover low and high gravity and vertical mixing, and solar and high-metallicity.} 
\label{fig:JWST_phot_cmd}
\end{center}
\end{figure*}

Color-magnitude diagrams provide a powerful means of distinguishing populations that differ in age, metallicity, atmospheric, and multiplicity properties \citep{2016ApJS..225...10F,2016ApJ...833...96L,2021ApJS..253....7K}. Figure~\ref{fig:JWST_phot_cmd} compares our low-gravity sample with field L and T dwarfs, directly imaged giant planets, and cloud-free Sonora Elf Owl atmosphere models. We adopt the same plotting conventions as in Figure~\ref{fig:JWST_phot}.

In all four panels, the field sequence traces the familiar L/T transition discussed in Section~\ref{subsec:colorabsvsspt}. Low-gravity objects are systematically overluminous and redder than field dwarfs of similar spectral type, extending the sequence toward brighter absolute magnitudes and redder colors. The VL-G objects define the most extreme locus, consistent with the trends identified in the color--spectral type and absolute magnitude--spectral type diagrams.

Most directly imaged giant planets closely follow the empirical low-gravity sequence. In particular, VHS~1256--1257~b, TWA~27~b, HR~8799~d, and $\beta$~Pic~b occupy the same region, or extend, the sequence of the reddest low-gravity brown dwarfs, suggesting broadly similar atmospheric properties. By contrast, HIP~65426~b, HR~8799~bce, and 29~Cyg~b are displaced toward the region occupied by models with enhanced metallicity or stronger vertical mixing. HR~8799~b is the most distinctive object, particularly in the (F410M--F460M) color where it is an extreme outlier, consistent with previous work \citep{Balmer2025}. At the opposite extreme, COCONUTS-2~b and $\beta$~Pic~d remain close to the field sequence and are broadly consistent with solar-metallicity, high-gravity models. This agrees with the old age of COCONUTS-2~b, and the cooler atmosphere of $\beta$~Pic~d \citep{Kiman2026,Gibbs2026}.

Taken together, the color--magnitude and color--color diagrams suggest that condensate clouds are the primary driver of the distinctive locus occupied by young, low-gravity objects. Enhanced cloud opacity redistributes flux from the near-infrared to longer wavelengths, producing both brighter absolute magnitudes and redder colors than observed for field dwarfs. The close agreement between VHS~1256--1257~b, TWA~27~b, HR~8799~d, $\beta$~Pic~b and the VL-G sequence suggests that these planets possess cloud properties similar to the youngest free-floating brown dwarfs. This interpretation is consistent with the spectroscopic analysis of \citet{Wang2022}, who found that HR~8799~d likely possesses more vertically extended and spatially uniform clouds than HR~8799~c. Our results further suggest that HR~8799~c more closely resembles HR~8799~e and 29~Cyg~b in its photometric properties.

The position of HR~8799~b is likely influenced by multiple atmospheric parameters. Its lower effective temperature produces stronger methane absorption than the other HR~8799 planets, while metallicity, vertical mixing, and cloud properties may also contribute to its distinctive colors \citep{Bonnefoy2016,Nasedkin2024}. More generally, these diagrams provide a useful empirical framework for interpreting populations of directly imaged planets. However, because several atmospheric parameters produce similar photometric signatures, robust characterization of individual objects ultimately requires spectroscopy and detailed atmospheric retrieval analyses.

\section{Conclusions}
\label{sec:conclusions}

In this work, we present the first empirical library of SPHEREx spectra for previously confirmed young, low-gravity L and T dwarfs and use it to investigate how low surface gravity shapes the $3$--$5\,{\rm \mu m}$ spectral energy distribution. We developed the open-source Python package \texttt{spherexminer} to extract spectra from the SPHEREx archive using aperture photometry. Given the large SPHEREx pixel scale, every extracted spectrum was visually inspected to ensure that it was free from contamination by nearby sources. Validation against SpeX prism spectra for a sample of field L and T dwarfs demonstrates that the extracted SPHEREx spectra accurately reproduce both the absolute flux calibration and overall spectral shape. This sample was adopted as the field comparison sample throughout this work.

We extracted SPHEREx spectra for 94 brown dwarfs and isolated planetary-mass objects with spectroscopic signatures of low gravity and youth from the UltracoolSheet. Using these spectra, we calculated synthetic JWST/NIRCam photometry and examined colors and absolute magnitudes as a function of spectral type. We find that colors spanning the near- and thermal-infrared, such as (F115W--F444W), provide powerful empirical diagnostics of low gravity. These colors are primarily sensitive to the redistribution of flux from the near-infrared to longer wavelengths caused by thick condensate clouds. In contrast, colors constructed from more closely spaced filters, particularly those confined to the $3$--$5\,{\rm \mu m}$ region, are more strongly influenced by molecular absorption and atmospheric chemistry.

Several thermal-infrared colors reveal additional atmospheric information. We find that (F410M--F460M) and ($L^\prime$--F444W) separate low-gravity and field objects despite spanning relatively narrow wavelength ranges. This behavior suggests differences in the apparent strengths of the CO and CO$_2$ absorption bands between the two populations, although atmospheric models do not predict substantially lower CO abundances in low-gravity atmospheres. Enhanced cloud opacity may reduce the apparent contrast of these molecular bands by shifting the photosphere to higher, cooler atmospheric layers, thus causing a similar effect in the (F410M--F460M) and ($L^\prime$--F444W) colors. In addition, colors such as (F335M--F444W) provide excellent tracers of spectral type by following the development of the fundamental CH$_4$ absorption band at $3.3\,{\rm \mu m}$.

Low-gravity objects are also systematically overluminous relative to field L dwarfs across all of the JWST filters examined in this work. This behavior is broadly consistent with the larger radii and higher bolometric luminosities expected for young substellar objects, although gravity-dependent spectral type--temperature relations and the effects of condensate clouds likely also contribute to the observed offsets.

Finally, we compared the empirical sequences defined by free-floating low-gravity objects with a sample of directly imaged giant planets, including HR~8799~bcde, $\beta$~Pic~bd, HIP~65426~b, 29~Cyg~b, TWA~27~b, VHS~1256--1257~b, and COCONUTS-2~b. Most directly imaged planets occupy the same locus as the youngest free-floating brown dwarfs in both the color--color and color--magnitude diagrams, suggesting broadly similar atmospheric properties. In particular, TWA~27~b, VHS~1256--1257~b, HR~8799~d, and $\beta$~Pic~b closely follow the empirical sequence of very low-gravity objects, consistent with thick condensate clouds dominating their atmospheric appearance. HR~8799~b is the principal exception, exhibiting colors consistent with its lower effective temperature and stronger methane absorption, while several planets occupy regions associated with enhanced metallicity or stronger vertical mixing in the Sonora Elf Owl models. COCONUTS-2~b and $\beta$~Pic~d remain consistent with the field sequence and solar-metallicity models. This agrees with the old age of COCONUTS-2~b, and the cold effective temperature of $\beta$~Pic~d. Although these diagrams provide valuable empirical diagnostics, multiple atmospheric parameters—including clouds, metallicity, chemistry, and vertical mixing—produce have similar photometric signatures. Detailed spectroscopic observations and atmospheric retrieval analyses remain essential for disentangling these effects in individual objects.

The empirical SPHEREx spectral library and JWST photometric sequences presented here provide a valuable observational reference for interpreting the atmospheres of young brown dwarfs and directly imaged giant planets. As SPHEREx continues to survey the sky and the sample of directly imaged exoplanets grows, these empirical benchmarks will become increasingly valuable for distinguishing the effects of surface gravity, clouds, atmospheric chemistry, and evolution in young substellar atmospheres.

\begin{acknowledgments}

RK would like to thank Davy Kirkpatrick, Zafar Rustamkulov and Federico Marocco for useful discussions about SPHEREx. 
J.W.X is grateful for support from the Heising-Simons Foundation 51 Pegasi b Fellowship (grant \#2025-5887).
This research has made use of the NASA/IPAC Infrared Science Archive, which is funded by the National Aeronautics and Space Administration and operated by the California Institute of Technology.
This publication makes use of data products from the Spectro-Photometer for the History of the Universe, Epoch of Reionization and Ices Explorer (SPHEREx), which is a joint project of the Jet Propulsion Laboratory and the California Institute of Technology, and is funded by the National Aeronautics and Space Administration.
This research made use of Photutils, an Astropy package for
detection and photometry of astronomical sources \citep{bradley_2026_19636730}.
This work has benefited from The UltracoolSheet at \url{http://bit.ly/UltracoolSheet}, maintained by Will Best, Trent Dupuy, Michael Liu, Aniket Sanghi, Rob Siverd, and Zhoujian Zhang, and developed from compilations by 
\citet{2012ApJS..201...19D,Dupuy2013,Deacon2014,2016ApJ...833...96L,Best2018,Best2021,2023ApJ...959...63S,Schneider2023}.	

\end{acknowledgments}

%
\facilities{IRSA, JWST, SPHEREx}

\software{
           \texttt{astropy} \citep{astropy2013,astropy2018,astropy2022}; 
           \texttt{astroquery} \citep{2019AJ....157...98G};
           \texttt{matplotlib} \citep{Hunter2007}; 
           \texttt{numpy} \citep{harris2020array}; 
           \texttt{photutils} \citep{bradley_2026_19636730};
           \texttt{scipy} \citep{SciPy-NMeth2020}; 
           \texttt{spherexminer} (This work)
           }





\bibliography{bibliography.bib,packages.bib}{}

@ARTICLE{Ackerman2001,
       author = {{Ackerman}, Andrew S. and {Marley}, Mark S.},
        title = "{Precipitating Condensation Clouds in Substellar Atmospheres}",
      journal = {\apj},
         year = 2001,
        month = aug,
       volume = {556},
       number = {2},
        pages = {872-884},
          doi = {10.1086/321540},
archivePrefix = {arXiv},
       eprint = {astro-ph/0103423},
 primaryClass = {astro-ph},
       adsurl = {https://ui.adsabs.harvard.edu/abs/2001ApJ...556..872A}
}

@ARTICLE{Akeson2025,
       author = {{Akeson}, Rachel and {Dubois-Felsmann}, Gregory P. and {Crill}, Brendan P. and {Faisst}, Andreas L. and {Fatahi}, Tamim and {Fazar}, Candice M. and {Goldina}, Tatiana and {Masters}, Daniel C. and {Nelson}, Christina and {Paladini}, Roberta and {Teplitz}, Harry I. and {Torrini}, Gabriela and {Velicheti}, Phani and {Ashby}, Matthew L.~N. and {Avner}, Dan and {Bach}, Yoonsoo P. and {Bock}, James J. and {Bruton}, Sean and {Bryan}, Sean A. and {Chang}, Tzu-Ching and {Chen}, Shuang-Shuang and {Cukierman}, Ari J. and {Dore}, O. and {Dowell}, C. Darren and {Everett}, Spencer and {Feder}, Richard M. and {Huai}, Zhaoyu and {Hui}, Howard and {Jeong}, Woong-Seob and {Jo}, Young-Soo and {Korngut}, Phil M. and {Kwon}, Yuna G. and {Lee}, Bomee and {Melnick}, Gary J. and {Murgia}, Giulia and {Nguyen}, Chi H. and {Pourrahmani}, Milad and {Rustamkulov}, Zafar and {Tolls}, Volker and {Wang}, Pao-Yu and {Yang}, Yujin and {Zemcov}, Michael},
        title = "{The SPHEREx Image and Spectrophotometry Processing Pipeline}",
      journal = {arXiv e-prints},
         year = 2025,
        month = nov,
          eid = {arXiv:2511.15823},
        pages = {arXiv:2511.15823},
          doi = {10.48550/arXiv.2511.15823},
archivePrefix = {arXiv},
       eprint = {2511.15823},
 primaryClass = {astro-ph.IM},
       adsurl = {https://ui.adsabs.harvard.edu/abs/2025arXiv251115823A}
}

@ARTICLE{Apai2013,
       author = {{Apai}, D{\'a}niel and {Radigan}, Jacqueline and {Buenzli}, Esther and {Burrows}, Adam and {Reid}, Iain Neill and {Jayawardhana}, Ray},
        title = "{HST Spectral Mapping of L/T Transition Brown Dwarfs Reveals Cloud Thickness Variations}",
      journal = {\apj},
         year = 2013,
        month = may,
       volume = {768},
       number = {2},
          eid = {121},
        pages = {121},
          doi = {10.1088/0004-637X/768/2/121},
archivePrefix = {arXiv},
       eprint = {1303.4151},
 primaryClass = {astro-ph.EP},
       adsurl = {https://ui.adsabs.harvard.edu/abs/2013ApJ...768..121A}
}

@ARTICLE{Barman2011,
       author = {{Barman}, Travis S. and {Macintosh}, Bruce and {Konopacky}, Quinn M. and {Marois}, Christian},
        title = "{Clouds and Chemistry in the Atmosphere of Extrasolar Planet HR8799b}",
      journal = {\apj},
         year = 2011,
        month = may,
       volume = {733},
       number = {1},
          eid = {65},
        pages = {65},
          doi = {10.1088/0004-637X/733/1/65},
archivePrefix = {arXiv},
       eprint = {1103.3895},
 primaryClass = {astro-ph.EP},
       adsurl = {https://ui.adsabs.harvard.edu/abs/2011ApJ...733...65B}
}

@ARTICLE{Beiler2024,
       author = {{Beiler}, Samuel A. and {Cushing}, Michael C. and {Kirkpatrick}, J. Davy and {Schneider}, Adam C. and {Mukherjee}, Sagnick and {Marley}, Mark S. and {Marocco}, Federico and {Smart}, Richard L.},
        title = "{Precise Bolometric Luminosities and Effective Temperatures of 23 Late-T and Y Dwarfs Obtained with JWST}",
      journal = {\apj},
         year = 2024,
        month = oct,
       volume = {973},
       number = {2},
          eid = {107},
        pages = {107},
          doi = {10.3847/1538-4357/ad6301},
archivePrefix = {arXiv},
       eprint = {2407.08518},
 primaryClass = {astro-ph.SR},
       adsurl = {https://ui.adsabs.harvard.edu/abs/2024ApJ...973..107B}
}

@ARTICLE{Biller2017,
       author = {{Biller}, Beth},
        title = "{The time domain for brown dwarfs and directly imaged giant exoplanets: the power of variability monitoring}",
      journal = {The Astronomical Review},
         year = 2017,
        month = jan,
       volume = {13},
       number = {1},
        pages = {1-27},
          doi = {10.1080/21672857.2017.1303105},
       adsurl = {https://ui.adsabs.harvard.edu/abs/2017AstRv..13....1B}
}

@ARTICLE{Bonnefoy2016,
       author = {{Bonnefoy}, M. and {Zurlo}, A. and {Baudino}, J.~L. and {Lucas}, P. and {Mesa}, D. and {Maire}, A.-L. and {Vigan}, A. and {Galicher}, R. and {Homeier}, D. and {Marocco}, F. and {Gratton}, R. and {Chauvin}, G. and {Allard}, F. and {Desidera}, S. and {Kasper}, M. and {Moutou}, C. and {Lagrange}, A.-M. and {Antichi}, J. and {Baruffolo}, A. and {Baudrand}, J. and {Beuzit}, J.-L. and {Boccaletti}, A. and {Cantalloube}, F. and {Carbillet}, M. and {Charton}, J. and {Claudi}, R.~U. and {Costille}, A. and {Dohlen}, K. and {Dominik}, C. and {Fantinel}, D. and {Feautrier}, P. and {Feldt}, M. and {Fusco}, T. and {Gigan}, P. and {Girard}, J.~H. and {Gluck}, L. and {Gry}, C. and {Henning}, T. and {Janson}, M. and {Langlois}, M. and {Madec}, F. and {Magnard}, Y. and {Maurel}, D. and {Mawet}, D. and {Meyer}, M.~R. and {Milli}, J. and {Moeller-Nilsson}, O. and {Mouillet}, D. and {Pavlov}, A. and {Perret}, D. and {Pujet}, P. and {Quanz}, S.~P. and {Rochat}, S. and {Rousset}, G. and {Roux}, A. and {Salasnich}, B. and {Salter}, G. and {Sauvage}, J.-F. and {Schmid}, H.~M. and {Sevin}, A. and {Soenke}, C. and {Stadler}, E. and {Turatto}, M. and {Udry}, S. and {Vakili}, F. and {Wahhaj}, Z. and {Wildi}, F.},
        title = "{First light of the VLT planet finder SPHERE. IV. Physical and chemical properties of the planets around HR8799}",
      journal = {\aap},
         year = 2016,
        month = mar,
       volume = {587},
          eid = {A58},
        pages = {A58},
          doi = {10.1051/0004-6361/201526906},
archivePrefix = {arXiv},
       eprint = {1511.04082},
 primaryClass = {astro-ph.EP},
       adsurl = {https://ui.adsabs.harvard.edu/abs/2016A&A...587A..58B}
}

@ARTICLE{Bowler2016,
       author = {{Bowler}, Brendan P.},
        title = "{Imaging Extrasolar Giant Planets}",
      journal = {\pasp},
         year = 2016,
        month = oct,
       volume = {128},
       number = {968},
        pages = {102001},
          doi = {10.1088/1538-3873/128/968/102001},
archivePrefix = {arXiv},
       eprint = {1605.02731},
 primaryClass = {astro-ph.EP},
       adsurl = {https://ui.adsabs.harvard.edu/abs/2016PASP..128j2001B}
}

@article{Bowler2010, 
year = {2010}, 
title = {{Near-infrared Spectroscopy of the Extrasolar Planet HR 8799 b}}, 
author = {Bowler, Brendan P and Liu, Michael C and Dupuy, Trent J and Cushing, Michael C}, 
journal = {The Astrophysical Journal}, 
doi = {10.1088/0004-637x/723/1/850}, 
url = {http://adsabs.harvard.edu/cgi-bin/nph-data\_query?bibcode=2010ApJ...723..850B\&link\_type=ABSTRACT}, 
pages = {850}, 
volume = {723}, 
month = {11}
}

@ARTICLE{Brooks2026,
       author = {{Brooks}, Hunter and {Cushing}, Michael C. and {Kothari}, Harshil},
        title = "{An Ultracool Dwarf Spectral Sequence Using SPHEREx}",
      journal = {Research Notes of the American Astronomical Society},
         year = 2026,
        month = apr,
       volume = {10},
       number = {4},
          eid = {94},
        pages = {94},
          doi = {10.3847/2515-5172/ae6257},
       adsurl = {https://ui.adsabs.harvard.edu/abs/2026RNAAS..10...94B}
}

@ARTICLE{Chauvin2004,
       author = {{Chauvin}, G. and {Lagrange}, A.-M. and {Dumas}, C. and {Zuckerman}, B. and {Mouillet}, D. and {Song}, I. and {Beuzit}, J.-L. and {Lowrance}, P.},
        title = "{A giant planet candidate near a young brown dwarf. Direct VLT/NACO observations using IR wavefront sensing}",
      journal = {\aap},
         year = 2004,
        month = oct,
       volume = {425},
        pages = {L29-L32},
          doi = {10.1051/0004-6361:200400056},
archivePrefix = {arXiv},
       eprint = {astro-ph/0409323},
 primaryClass = {astro-ph},
       adsurl = {https://ui.adsabs.harvard.edu/abs/2004A&A...425L..29C}
}

@ARTICLE{Currie2023,
       author = {{Currie}, Thayne and {Brandt}, G. Mirek and {Brandt}, Timothy D. and {Lacy}, Brianna and {Burrows}, Adam and {Guyon}, Olivier and {Tamura}, Motohide and {Liu}, Ranger Y. and {Sagynbayeva}, Sabina and {Tobin}, Taylor and {Chilcote}, Jeffrey and {Groff}, Tyler and {Marois}, Christian and {Thompson}, William and {Murphy}, Simon J. and {Kuzuhara}, Masayuki and {Lawson}, Kellen and {Lozi}, Julien and {Deo}, Vincent and {Vievard}, Sebastien and {Skaf}, Nour and {Uyama}, Taichi and {Jovanovic}, Nemanja and {Martinache}, Frantz and {Kasdin}, N. Jeremy and {Kudo}, Tomoyuki and {McElwain}, Michael and {Janson}, Markus and {Wisniewski}, John and {Hodapp}, Klaus and {Nishikawa}, Jun and {He{\l}miniak}, Krzysztof and {Kwon}, Jungmi and {Hayashi}, Masahiko},
        title = "{Direct imaging and astrometric detection of a gas giant planet orbiting an accelerating star}",
      journal = {Science},
         year = 2023,
        month = apr,
       volume = {380},
       number = {6641},
        pages = {198-203},
          doi = {10.1126/science.abo6192},
archivePrefix = {arXiv},
       eprint = {2212.00034},
 primaryClass = {astro-ph.EP},
       adsurl = {https://ui.adsabs.harvard.edu/abs/2023Sci...380..198C}
}

@ARTICLE{Malin2025,
       author = {{M{\^a}lin}, Mathilde and {Boccaletti}, Anthony and {Perrot}, Cl{\'e}ment and {Baudoz}, Pierre and {Rouan}, Daniel and {Lagage}, Pierre-Olivier and {Waters}, Rens and {G{\"u}del}, Manuel and {Henning}, Thomas and {Vandenbussche}, Bart and {Absil}, Olivier and {Barrado}, David and {Charnay}, Benjamin and {Choquet}, Elodie and {Cossou}, Christophe and {Danielski}, Camilla and {Decin}, Leen and {Glauser}, Adrian M. and {Pye}, John and {Olofsson}, Goran and {Glasse}, Alistair and {Patapis}, Polychronis and {Royer}, Pierre and {Scheithauer}, Silvia and {Serabyn}, Eugene and {Tremblin}, Pascal and {Whiteford}, Niall and {van Dishoeck}, Ewine F. and {Ostlin}, G{\"o}ran and {Ray}, Tom P. and {Wright}, Gillian},
        title = "{First unambiguous detection of ammonia in the atmosphere of a planetary mass companion with JWST/MIRI coronagraphs}",
      journal = {\aap},
         year = 2025,
        month = jan,
       volume = {693},
          eid = {A315},
        pages = {A315},
          doi = {10.1051/0004-6361/202452695},
archivePrefix = {arXiv},
       eprint = {2501.00104},
 primaryClass = {astro-ph.EP},
       adsurl = {https://ui.adsabs.harvard.edu/abs/2025A&A...693A.315M}
}

@ARTICLE{Marley2012,
       author = {{Marley}, Mark S. and {Saumon}, Didier and {Cushing}, Michael and {Ackerman}, Andrew S. and {Fortney}, Jonathan J. and {Freedman}, Richard},
        title = "{Masses, Radii, and Cloud Properties of the HR 8799 Planets}",
      journal = {\apj},
         year = 2012,
        month = aug,
       volume = {754},
       number = {2},
          eid = {135},
        pages = {135},
          doi = {10.1088/0004-637X/754/2/135},
archivePrefix = {arXiv},
       eprint = {1205.6488},
 primaryClass = {astro-ph.EP},
       adsurl = {https://ui.adsabs.harvard.edu/abs/2012ApJ...754..135M}
}

@ARTICLE{Marley2015,
       author = {{Marley}, M.~S. and {Robinson}, T.~D.},
        title = "{On the Cool Side: Modeling the Atmospheres of Brown Dwarfs and Giant Planets}",
      journal = {\araa},
         year = 2015,
        month = aug,
       volume = {53},
        pages = {279-323},
          doi = {10.1146/annurev-astro-082214-122522},
archivePrefix = {arXiv},
       eprint = {1410.6512},
 primaryClass = {astro-ph.EP},
       adsurl = {https://ui.adsabs.harvard.edu/abs/2015ARA&A..53..279M}
}

@ARTICLE{Nasedkin2024,
       author = {{Nasedkin}, E. and {Molli{\`e}re}, P. and {Lacour}, S. and {Nowak}, M. and {Kreidberg}, L. and {Stolker}, T. and {Wang}, J.~J. and {Balmer}, W.~O. and {Kammerer}, J. and {Shangguan}, J. and {Abuter}, R. and {Amorim}, A. and {Asensio-Torres}, R. and {Benisty}, M. and {Berger}, J.-P. and {Beust}, H. and {Blunt}, S. and {Boccaletti}, A. and {Bonnefoy}, M. and {Bonnet}, H. and {Bordoni}, M.~S. and {Bourdarot}, G. and {Brandner}, W. and {Cantalloube}, F. and {Caselli}, P. and {Charnay}, B. and {Chauvin}, G. and {Chavez}, A. and {Choquet}, E. and {Christiaens}, V. and {Cl{\'e}net}, Y. and {Coud{\'e} Du Foresto}, V. and {Cridland}, A. and {Davies}, R. and {Dembet}, R. and {Dexter}, J. and {Drescher}, A. and {Duvert}, G. and {Eckart}, A. and {Eisenhauer}, F. and {F{\"o}rster Schreiber}, N.~M. and {Garcia}, P. and {Garcia Lopez}, R. and {Gendron}, E. and {Genzel}, R. and {Gillessen}, S. and {Girard}, J.~H. and {Grant}, S. and {Haubois}, X. and {Hei{\ss}el}, G. and {Henning}, Th. and {Hinkley}, S. and {Hippler}, S. and {Houll{\'e}}, M. and {Hubert}, Z. and {Jocou}, L. and {Keppler}, M. and {Kervella}, P. and {Kurtovic}, N.~T. and {Lagrange}, A.-M. and {Lapeyr{\`e}re}, V. and {Le Bouquin}, J.-B. and {Lutz}, D. and {Maire}, A.-L. and {Mang}, F. and {Marleau}, G.-D. and {M{\'e}rand}, A. and {Monnier}, J.~D. and {Mordasini}, C. and {Ott}, T. and {Otten}, G.~P.~P.~L. and {Paladini}, C. and {Paumard}, T. and {Perraut}, K. and {Perrin}, G. and {Pfuhl}, O. and {Pourr{\'e}}, N. and {Pueyo}, L. and {Ribeiro}, D.~C. and {Rickman}, E. and {Ruffio}, J.~B. and {Rustamkulov}, Z. and {Shimizu}, T. and {Sing}, D. and {Stadler}, J. and {Straub}, O. and {Straubmeier}, C. and {Sturm}, E. and {Tacconi}, L.~J. and {van Dishoeck}, E.~F. and {Vigan}, A. and {Vincent}, F. and {von Fellenberg}, S.~D. and {Widmann}, F. and {Winterhalder}, T.~O. and {Woillez}, J. and {Yazici}, {\c{S}}. and {Gravity Collaboration}},
        title = "{Four-of-a-kind? Comprehensive atmospheric characterisation of the HR 8799 planets with VLTI/GRAVITY}",
      journal = {\aap},
         year = 2024,
        month = jul,
       volume = {687},
          eid = {A298},
        pages = {A298},
          doi = {10.1051/0004-6361/202449328},
archivePrefix = {arXiv},
       eprint = {2404.03776},
 primaryClass = {astro-ph.EP},
       adsurl = {https://ui.adsabs.harvard.edu/abs/2024A&A...687A.298N}
}

@ARTICLE{2013ApJ...772...79A,
       author = {{Allers}, K.~N. and {Liu}, Michael C.},
        title = "{A Near-infrared Spectroscopic Study of Young Field Ultracool Dwarfs}",
      journal = {\apj},
         year = 2013,
        month = aug,
       volume = {772},
       number = {2},
          eid = {79},
        pages = {79},
          doi = {10.1088/0004-637X/772/2/79},
archivePrefix = {arXiv},
       eprint = {1305.4418},
 primaryClass = {astro-ph.SR},
       adsurl = {https://ui.adsabs.harvard.edu/abs/2013ApJ...772...79A}
}

@ARTICLE{Balmer2025,
       author = {{Balmer}, William O. and {Kammerer}, Jens and {Pueyo}, Laurent and {Perrin}, Marshall D. and {Girard}, Julien H. and {Leisenring}, Jarron M. and {Lawson}, Kellen and {Dennen}, Henry and {van der Marel}, Roeland P. and {Beichman}, Charles A. and {Bryden}, Geoffrey and {Llop-Sayson}, Jorge and {Valenti}, Jeff A. and {Lothringer}, Joshua D. and {Lewis}, Nikole K. and {M{\^a}lin}, Mathilde and {Rebollido}, Isabel and {Rickman}, Emily and {Hoch}, Kielan K.~W. and {Soummer}, R{\'e}mi and {Clampin}, Mark and {Mountain}, C. Matt},
        title = "{JWST-TST High Contrast: Living on the Wedge, or, NIRCam Bar Coronagraphy Reveals CO$_{2}$ in the HR 8799 and 51 Eri Exoplanets' Atmospheres}",
      journal = {\aj},
         year = 2025,
        month = apr,
       volume = {169},
       number = {4},
          eid = {209},
        pages = {209},
          doi = {10.3847/1538-3881/adb1c6},
archivePrefix = {arXiv},
       eprint = {2503.13608},
 primaryClass = {astro-ph.EP},
       adsurl = {https://ui.adsabs.harvard.edu/abs/2025AJ....169..209B}
}

@ARTICLE{Barman2015,
       author = {{Barman}, Travis S. and {Konopacky}, Quinn M. and {Macintosh}, Bruce and {Marois}, Christian},
        title = "{Simultaneous Detection of Water, Methane, and Carbon Monoxide in the Atmosphere of Exoplanet HR8799b}",
      journal = {\apj},
         year = 2015,
        month = may,
       volume = {804},
       number = {1},
          eid = {61},
        pages = {61},
          doi = {10.1088/0004-637X/804/1/61},
archivePrefix = {arXiv},
       eprint = {1503.03539},
 primaryClass = {astro-ph.EP},
       adsurl = {https://ui.adsabs.harvard.edu/abs/2015ApJ...804...61B}
}

@ARTICLE{Xuan2024,
       author = {{Xuan}, Jerry W. and {Hsu}, Chih-Chun and {Finnerty}, Luke and {Wang}, Jason and {Ruffio}, Jean-Baptiste and {Zhang}, Yapeng and {Knutson}, Heather A. and {Mawet}, Dimitri and {Mamajek}, Eric E. and {Inglis}, Julie and {Wallack}, Nicole L. and {Bryan}, Marta L. and {Blake}, Geoffrey A. and {Molli{\`e}re}, Paul and {Hejazi}, Neda and {Baker}, Ashley and {Bartos}, Randall and {Calvin}, Benjamin and {Cetre}, Sylvain and {Delorme}, Jacques-Robert and {Doppmann}, Greg and {Echeverri}, Daniel and {Fitzgerald}, Michael P. and {Jovanovic}, Nemanja and {Liberman}, Joshua and {L{\'o}pez}, Ronald A. and {Morris}, Evan and {Pezzato}, Jacklyn and {Sappey}, Ben and {Schofield}, Tobias and {Skemer}, Andrew and {Wallace}, J. Kent and {Wang}, Ji and {Agrawal}, Shubh and {Horstman}, Katelyn},
        title = "{Are These Planets or Brown Dwarfs? Broadly Solar Compositions from High-resolution Atmospheric Retrievals of {\ensuremath{\sim}}10{\textendash}30 M $_{Jup}$ Companions}",
      journal = {\apj},
         year = 2024,
        month = jul,
       volume = {970},
       number = {1},
          eid = {71},
        pages = {71},
          doi = {10.3847/1538-4357/ad4796},
archivePrefix = {arXiv},
       eprint = {2405.13128},
 primaryClass = {astro-ph.EP},
       adsurl = {https://ui.adsabs.harvard.edu/abs/2024ApJ...970...71X}
}

@ARTICLE{Molliere2020,
       author = {{Molli{\`e}re}, P. and {Stolker}, T. and {Lacour}, S. and {Otten}, G.~P.~P.~L. and {Shangguan}, J. and {Charnay}, B. and {Molyarova}, T. and {Nowak}, M. and {Henning}, Th. and {Marleau}, G.-D. and {Semenov}, D.~A. and {van Dishoeck}, E. and {Eisenhauer}, F. and {Garcia}, P. and {Garcia Lopez}, R. and {Girard}, J.~H. and {Greenbaum}, A.~Z. and {Hinkley}, S. and {Kervella}, P. and {Kreidberg}, L. and {Maire}, A.-L. and {Nasedkin}, E. and {Pueyo}, L. and {Snellen}, I.~A.~G. and {Vigan}, A. and {Wang}, J. and {de Zeeuw}, P.~T. and {Zurlo}, A.},
        title = "{Retrieving scattering clouds and disequilibrium chemistry in the atmosphere of HR 8799e}",
      journal = {\aap},
         year = 2020,
        month = aug,
       volume = {640},
          eid = {A131},
        pages = {A131},
          doi = {10.1051/0004-6361/202038325},
archivePrefix = {arXiv},
       eprint = {2006.09394},
 primaryClass = {astro-ph.EP},
       adsurl = {https://ui.adsabs.harvard.edu/abs/2020A&A...640A.131M}
}

@ARTICLE{Balmer2026,
       author = {{Balmer}, William O. and {Pueyo}, Laurent and {Messier}, Ashley and {Bruinsma}, Evelyn and {Jones}, Jeremy and {Matuszewska}, Klara and {Perrin}, Marshall D. and {Girard}, Julien H. and {Leisenring}, Jarron M. and {Lawson}, Kellen and {van der Marel}, Roeland P. and {Kammerer}, Jens and {Carter}, Aarynn and {M{\^a}lin}, Mathilde and {Ward-Duong}, Kimberly and {Hoch}, Kielan K.~W. and {Rickman}, Emily and {Seager}, Sara},
        title = "{Direct Images of CO$_{2}$ Absorption in the Atmosphere of a Super-Jupiter: Enhanced Metallicity Suggestive of Formation in a Disk}",
      journal = {\apjl},
         year = 2026,
        month = apr,
       volume = {1001},
       number = {2},
          eid = {L26},
        pages = {L26},
          doi = {10.3847/2041-8213/ae374a},
archivePrefix = {arXiv},
       eprint = {2604.09785},
 primaryClass = {astro-ph.EP},
       adsurl = {https://ui.adsabs.harvard.edu/abs/2026ApJ..1001L..26B}
}

@ARTICLE{Best2018,
       author = {{Best}, William M.~J. and {Magnier}, Eugene A. and {Liu}, Michael C. and {Aller}, Kimberly M. and {Zhang}, Zhoujian and {Burgett}, W.~S. and {Chambers}, K.~C. and {Draper}, P. and {Flewelling}, H. and {Kaiser}, N. and {Kudritzki}, R.-P. and {Metcalfe}, N. and {Tonry}, J.~L. and {Wainscoat}, R.~J. and {Waters}, C.},
        title = "{Photometry and Proper Motions of M, L, and T Dwarfs from the Pan-STARRS1 3{\ensuremath{\pi}} Survey}",
      journal = {\apjs},
         year = 2018,
        month = jan,
       volume = {234},
       number = {1},
          eid = {1},
        pages = {1},
          doi = {10.3847/1538-4365/aa9982},
archivePrefix = {arXiv},
       eprint = {1701.00490},
 primaryClass = {astro-ph.SR},
       adsurl = {https://ui.adsabs.harvard.edu/abs/2018ApJS..234....1B}
}

@ARTICLE{Best2021,
       author = {{Best}, William M.~J. and {Liu}, Michael C. and {Magnier}, Eugene A. and {Dupuy}, Trent J.},
        title = "{A Volume-limited Sample of Ultracool Dwarfs. I. Construction, Space Density, and a Gap in the L/T Transition}",
      journal = {\aj},
         year = 2021,
        month = jan,
       volume = {161},
       number = {1},
          eid = {42},
        pages = {42},
          doi = {10.3847/1538-3881/abc893},
archivePrefix = {arXiv},
       eprint = {2010.15853},
 primaryClass = {astro-ph.SR},
       adsurl = {https://ui.adsabs.harvard.edu/abs/2021AJ....161...42B}
}

@ARTICLE{Bock2026,
       author = {{Bock}, James J. and {Aboobaker}, Asad M. and {Adamo}, Joseph and {Akeson}, Rachel and {Alred}, John M. and {Alibay}, Farah and {Ashby}, Matthew L.~N. and {Bach}, Yoonsoo P. and {Bleem}, Lindsey E. and {Bolton}, Douglas and {Braun}, David F. and {Bruton}, Sean and {Bryan}, Sean A. and {Chang}, Tzu-Ching and {Chen}, Shuang-Shuang and {Cheng}, Yun-Ting and {Cheshire}, IV, James R. and {Chiang}, Yi-Kuan and {Choppin de Janvry}, Jean and {Condon}, Samuel and {Cook}, Walter R. and {Cooray}, Asantha and {Crill}, Brendan P. and {Cukierman}, Ari J. and {Dor{\'e}}, Olivier and {Dowell}, C. Darren and {Dubois-Felsmann}, Gregory P. and {Eifler}, Tim and {Everett}, Spencer and {Fabinsky}, Beth E. and {Faisst}, Andreas L. and {Fanson}, James L. and {Farrington}, Allen H. and {Fatahi}, Tamim and {Fazar}, Candice M. and {Feder}, Richard M. and {Frater}, Eric H. and {Grasshorn Gebhardt}, Henry S. and {Giri}, Utkarsh and {Goldina}, Tatiana and {Gorjian}, Varoujan and {Habib}, Salman and {Hart}, William G. and {Heinrich}, Chen and {Hora}, Joseph L. and {Huai}, Zhaoyu and {Hui}, Howard and {Jo}, Young-Soo and {Jeong}, Woong-Seob and {Kang}, Jae Hwan and {Kang}, Miju and {Kecman}, Branislav and {Kim}, Chul-Hwan and {Kim}, Jaeyeong and {Kim}, Minjin and {Kim}, Young-Jun and {Kim}, Yongjung and {Kirkpatrick}, J. Davy and {Kobayashi}, Yosuke and {Korngut}, Phil M. and {Krause}, Elisabeth and {Lee}, Bomee and {Lee}, Ho-Gyu and {Lee}, Jae-Joon and {Lee}, Jeong-Eun and {Lisse}, Carey M. and {Mariani}, Giacomo and {Masters}, Daniel C. and {Mauskopf}, Philip D. and {Melnick}, Gary J. and {Minasyan}, Mary H. and {Mirocha}, Jordan and {Miyasaka}, Hiromasa and {Moore}, Anne and {Moore}, Bradley D. and {Murgia}, Giulia and {Naylor}, Bret J. and {Nelson}, Christina and {Nguyen}, Chi H. and {Nguyen}, Hien T. and {Noh}, Jinyoung K. and {Padin}, Stephen and {Paladini}, Roberta and {Park}, Sung-Joon and {Penanen}, Konstantin I. and {Putnam}, Dustin S. and {Pyo}, Jeonghyun and {Ramachandra}, Nesar and {Ramanathan}, Keshav and {Rustamkulov}, Zafar and {Reiley}, Daniel J. and {Rice}, Eric B. and {Rocca}, Jennifer M. and {Seok}, Ji Yeon and {Smith}, Roger and {Stober}, Jeremy and {Susca}, Sara and {Teplitz}, Harry I. and {Thelen}, Michael P. and {Tolls}, Volker and {Torrini}, Gabriela and {Trangsrud}, Amy R. and {Unwin}, Stephen and {Velicheti}, Phani and {Wang}, Pao-Yu and {Wen}, Robin Y. and {Werner}, Michael W. and {Williams}, Abby E. and {Williamson}, Ross and {Wincentsen}, James and {Windhorst}, Rogier A. and {Yang}, Soung-Chul and {Yang}, Yujin and {Zemcov}, Michael},
        title = "{The SPHEREx Satellite Mission}",
      journal = {\apj},
         year = 2026,
        month = mar,
       volume = {999},
       number = {1},
          eid = {139},
        pages = {139},
          doi = {10.3847/1538-4357/ae2be2},
archivePrefix = {arXiv},
       eprint = {2511.02985},
 primaryClass = {astro-ph.IM},
       adsurl = {https://ui.adsabs.harvard.edu/abs/2026ApJ...999..139B}
}

@ARTICLE{Bowler2014,
       author = {{Bowler}, Brendan P. and {Liu}, Michael C. and {Kraus}, Adam L. and {Mann}, Andrew W.},
        title = "{Spectroscopic Confirmation of Young Planetary-mass Companions on Wide Orbits}",
      journal = {\apj},
         year = 2014,
        month = mar,
       volume = {784},
       number = {1},
          eid = {65},
        pages = {65},
          doi = {10.1088/0004-637X/784/1/65},
archivePrefix = {arXiv},
       eprint = {1401.7668},
 primaryClass = {astro-ph.EP},
       adsurl = {https://ui.adsabs.harvard.edu/abs/2014ApJ...784...65B}
}

@ARTICLE{Burgasser2002,
       author = {{Burgasser}, Adam J. and {Marley}, Mark S. and {Ackerman}, Andrew S. and {Saumon}, Didier and {Lodders}, Katharina and {Dahn}, Conard C. and {Harris}, Hugh C. and {Kirkpatrick}, J. Davy},
        title = "{Evidence of Cloud Disruption in the L/T Dwarf Transition}",
      journal = {\apjl},
         year = 2002,
        month = jun,
       volume = {571},
       number = {2},
        pages = {L151-L154},
          doi = {10.1086/341343},
archivePrefix = {arXiv},
       eprint = {astro-ph/0205051},
 primaryClass = {astro-ph},
       adsurl = {https://ui.adsabs.harvard.edu/abs/2002ApJ...571L.151B}
}

@ARTICLE{Burgasser2004,
       author = {{Burgasser}, Adam J. and {McElwain}, Michael W. and {Kirkpatrick}, J. Davy and {Cruz}, Kelle L. and {Tinney}, Chris G. and {Reid}, I. Neill},
        title = "{The 2MASS Wide-Field T Dwarf Search. III. Seven New T Dwarfs and Other Cool Dwarf Discoveries}",
      journal = {\aj},
         year = 2004,
        month = may,
       volume = {127},
       number = {5},
        pages = {2856-2870},
          doi = {10.1086/383549},
archivePrefix = {arXiv},
       eprint = {astro-ph/0402325},
 primaryClass = {astro-ph},
       adsurl = {https://ui.adsabs.harvard.edu/abs/2004AJ....127.2856B}
}

@ARTICLE{Burgasser2006,
       author = {{Burgasser}, Adam J. and {McElwain}, Michael W.},
        title = "{Resolved Spectroscopy of M Dwarf/L Dwarf Binaries. I. DENIS J220002.05-303832.9AB}",
      journal = {\aj},
         year = 2006,
        month = feb,
       volume = {131},
       number = {2},
        pages = {1007-1014},
          doi = {10.1086/499042},
archivePrefix = {arXiv},
       eprint = {astro-ph/0510579},
 primaryClass = {astro-ph},
       adsurl = {https://ui.adsabs.harvard.edu/abs/2006AJ....131.1007B}
}

@ARTICLE{2006ApJ...637.1067B,
       author = {{Burgasser}, Adam J. and {Geballe}, T.~R. and {Leggett}, S.~K. and {Kirkpatrick}, J. Davy and {Golimowski}, David A.},
        title = "{A Unified Near-Infrared Spectral Classification Scheme for T Dwarfs}",
      journal = {\apj},
         year = 2006,
        month = feb,
       volume = {637},
       number = {2},
        pages = {1067-1093},
          doi = {10.1086/498563},
archivePrefix = {arXiv},
       eprint = {astro-ph/0510090},
 primaryClass = {astro-ph},
       adsurl = {https://ui.adsabs.harvard.edu/abs/2006ApJ...637.1067B}
}

@ARTICLE{Burgasser2006c,
       author = {{Burgasser}, Adam J. and {Burrows}, Adam and {Kirkpatrick}, J. Davy},
        title = "{A Method for Determining the Physical Properties of the Coldest Known Brown Dwarfs}",
      journal = {\apj},
         year = 2006,
        month = mar,
       volume = {639},
       number = {2},
        pages = {1095-1113},
          doi = {10.1086/499344},
archivePrefix = {arXiv},
       eprint = {astro-ph/0510707},
 primaryClass = {astro-ph},
       adsurl = {https://ui.adsabs.harvard.edu/abs/2006ApJ...639.1095B}
}

@ARTICLE{Burgasser2007,
       author = {{Burgasser}, Adam J. and {Looper}, Dagny L. and {Kirkpatrick}, J. Davy and {Liu}, Michael C.},
        title = "{Discovery of a High Proper Motion L Dwarf Binary: 2MASS J15200224-4422419AB}",
      journal = {\apj},
         year = 2007,
        month = mar,
       volume = {658},
       number = {1},
        pages = {557-568},
          doi = {10.1086/511518},
archivePrefix = {arXiv},
       eprint = {astro-ph/0611697},
 primaryClass = {astro-ph},
       adsurl = {https://ui.adsabs.harvard.edu/abs/2007ApJ...658..557B}
}

@ARTICLE{Burgasser2007b,
       author = {{Burgasser}, Adam J.},
        title = "{Binaries and the L Dwarf/T Dwarf Transition}",
      journal = {\apj},
         year = 2007,
        month = apr,
       volume = {659},
       number = {1},
        pages = {655-674},
          doi = {10.1086/511027},
archivePrefix = {arXiv},
       eprint = {astro-ph/0611505},
 primaryClass = {astro-ph},
       adsurl = {https://ui.adsabs.harvard.edu/abs/2007ApJ...659..655B}
}

@ARTICLE{Burgasser2008,
       author = {{Burgasser}, Adam J. and {Liu}, Michael C. and {Ireland}, Michael J. and {Cruz}, Kelle L. and {Dupuy}, Trent J.},
        title = "{Subtle Signatures of Multiplicity in Late-type Dwarf Spectra: The Unresolved M8.5 + T5 Binary 2MASS J03202839-0446358}",
      journal = {\apj},
         year = 2008,
        month = jul,
       volume = {681},
       number = {1},
        pages = {579-593},
          doi = {10.1086/588379},
archivePrefix = {arXiv},
       eprint = {0803.0295},
 primaryClass = {astro-ph},
       adsurl = {https://ui.adsabs.harvard.edu/abs/2008ApJ...681..579B}
}

@ARTICLE{Burgasser2010,
       author = {{Burgasser}, Adam J. and {Cruz}, Kelle L. and {Cushing}, Michael and {Gelino}, Christopher R. and {Looper}, Dagny L. and {Faherty}, Jacqueline K. and {Kirkpatrick}, J. Davy and {Reid}, I. Neill},
        title = "{SpeX Spectroscopy of Unresolved Very Low Mass Binaries. I. Identification of 17 Candidate Binaries Straddling the L Dwarf/T Dwarf Transition}",
      journal = {\apj},
         year = 2010,
        month = feb,
       volume = {710},
       number = {2},
        pages = {1142-1169},
          doi = {10.1088/0004-637X/710/2/1142},
archivePrefix = {arXiv},
       eprint = {0912.3808},
 primaryClass = {astro-ph.SR},
       adsurl = {https://ui.adsabs.harvard.edu/abs/2010ApJ...710.1142B}
}

@ARTICLE{Burrows2001,
       author = {{Burrows}, Adam and {Hubbard}, W.~B. and {Lunine}, J.~I. and {Liebert}, James},
        title = "{The theory of brown dwarfs and extrasolar giant planets}",
      journal = {Reviews of Modern Physics},
         year = 2001,
        month = jul,
       volume = {73},
       number = {3},
        pages = {719-765},
          doi = {10.1103/RevModPhys.73.719},
archivePrefix = {arXiv},
       eprint = {astro-ph/0103383},
 primaryClass = {astro-ph},
       adsurl = {https://ui.adsabs.harvard.edu/abs/2001RvMP...73..719B}
}

@ARTICLE{Burrows2006,
       author = {{Burrows}, Adam and {Sudarsky}, David and {Hubeny}, Ivan},
        title = "{L and T Dwarf Models and the L to T Transition}",
      journal = {\apj},
         year = 2006,
        month = apr,
       volume = {640},
       number = {2},
        pages = {1063-1077},
          doi = {10.1086/500293},
archivePrefix = {arXiv},
       eprint = {astro-ph/0509066},
 primaryClass = {astro-ph},
       adsurl = {https://ui.adsabs.harvard.edu/abs/2006ApJ...640.1063B}
}

@ARTICLE{Chiu2006,
       author = {{Chiu}, K. and {Fan}, X. and {Leggett}, S.~K. and {Golimowski}, D.~A. and {Zheng}, W. and {Geballe}, T.~R. and {Schneider}, D.~P. and {Brinkmann}, J.},
        title = "{Seventy-One New L and T Dwarfs from the Sloan Digital Sky Survey}",
      journal = {\aj},
         year = 2006,
        month = jun,
       volume = {131},
       number = {5},
        pages = {2722-2736},
          doi = {10.1086/501431},
archivePrefix = {arXiv},
       eprint = {astro-ph/0601089},
 primaryClass = {astro-ph},
       adsurl = {https://ui.adsabs.harvard.edu/abs/2006AJ....131.2722C}
}

@ARTICLE{Cruz2009,
       author = {{Cruz}, Kelle L. and {Kirkpatrick}, J. Davy and {Burgasser}, Adam J.},
        title = "{Young L Dwarfs Identified in the Field: A Preliminary Low-Gravity, Optical Spectral Sequence from L0 to L5}",
      journal = {\aj},
         year = 2009,
        month = feb,
       volume = {137},
       number = {2},
        pages = {3345-3357},
          doi = {10.1088/0004-6256/137/2/3345},
archivePrefix = {arXiv},
       eprint = {0812.0364},
 primaryClass = {astro-ph},
       adsurl = {https://ui.adsabs.harvard.edu/abs/2009AJ....137.3345C}
}

@ARTICLE{Cushing2005,
       author = {{Cushing}, Michael C. and {Rayner}, John T. and {Vacca}, William D.},
        title = "{An Infrared Spectroscopic Sequence of M, L, and T Dwarfs}",
      journal = {\apj},
         year = 2005,
        month = apr,
       volume = {623},
       number = {2},
        pages = {1115-1140},
          doi = {10.1086/428040},
archivePrefix = {arXiv},
       eprint = {astro-ph/0412313},
 primaryClass = {astro-ph},
       adsurl = {https://ui.adsabs.harvard.edu/abs/2005ApJ...623.1115C}
}

@ARTICLE{Deacon2014,
       author = {{Deacon}, Niall R. and {Liu}, Michael C. and {Magnier}, Eugene A. and {Aller}, Kimberly M. and {Best}, William M.~J. and {Dupuy}, Trent and {Bowler}, Brendan P. and {Mann}, Andrew W. and {Redstone}, Joshua A. and {Burgett}, William S. and {Chambers}, Kenneth C. and {Draper}, Peter W. and {Flewelling}, H. and {Hodapp}, Klaus W. and {Kaiser}, Nick and {Kudritzki}, Rolf-Peter and {Morgan}, Jeff S. and {Metcalfe}, Nigel and {Price}, Paul A. and {Tonry}, John L. and {Wainscoat}, Richard J.},
        title = "{Wide Cool and Ultracool Companions to Nearby Stars from Pan-STARRS 1}",
      journal = {\apj},
         year = 2014,
        month = sep,
       volume = {792},
       number = {2},
          eid = {119},
        pages = {119},
          doi = {10.1088/0004-637X/792/2/119},
archivePrefix = {arXiv},
       eprint = {1407.2938},
 primaryClass = {astro-ph.SR},
       adsurl = {https://ui.adsabs.harvard.edu/abs/2014ApJ...792..119D}
}

@ARTICLE{2012ApJS..201...19D,
       author = {{Dupuy}, Trent J. and {Liu}, Michael C.},
        title = "{The Hawaii Infrared Parallax Program. I. Ultracool Binaries and the L/T Transition}",
      journal = {\apjs},
         year = 2012,
        month = aug,
       volume = {201},
       number = {2},
          eid = {19},
        pages = {19},
          doi = {10.1088/0067-0049/201/2/19},
archivePrefix = {arXiv},
       eprint = {1201.2465},
 primaryClass = {astro-ph.SR},
       adsurl = {https://ui.adsabs.harvard.edu/abs/2012ApJS..201...19D}
}

@ARTICLE{Dupuy2013,
       author = {{Dupuy}, Trent J. and {Kraus}, Adam L.},
        title = "{Distances, Luminosities, and Temperatures of the Coldest Known Substellar Objects}",
      journal = {Science},
         year = 2013,
        month = sep,
       volume = {341},
       number = {6153},
        pages = {1492-1495},
          doi = {10.1126/science.1241917},
archivePrefix = {arXiv},
       eprint = {1309.1422},
 primaryClass = {astro-ph.SR},
       adsurl = {https://ui.adsabs.harvard.edu/abs/2013Sci...341.1492D}
}

@ARTICLE{2016ApJS..225...10F,
       author = {{Faherty}, Jacqueline K. and {Riedel}, Adric R. and {Cruz}, Kelle L. and {Gagne}, Jonathan and {Filippazzo}, Joseph C. and {Lambrides}, Erini and {Fica}, Haley and {Weinberger}, Alycia and {Thorstensen}, John R. and {Tinney}, C.~G. and {Baldassare}, Vivienne and {Lemonier}, Emily and {Rice}, Emily L.},
        title = "{Population Properties of Brown Dwarf Analogs to Exoplanets}",
      journal = {\apjs},
         year = 2016,
        month = jul,
       volume = {225},
       number = {1},
          eid = {10},
        pages = {10},
          doi = {10.3847/0067-0049/225/1/10},
archivePrefix = {arXiv},
       eprint = {1605.07927},
 primaryClass = {astro-ph.SR},
       adsurl = {https://ui.adsabs.harvard.edu/abs/2016ApJS..225...10F}
}

@ARTICLE{Filippazzo2015,
       author = {{Filippazzo}, Joseph C. and {Rice}, Emily L. and {Faherty}, Jacqueline and {Cruz}, Kelle L. and {Van Gordon}, Mollie M. and {Looper}, Dagny L.},
        title = "{Fundamental Parameters and Spectral Energy Distributions of Young and Field Age Objects with Masses Spanning the Stellar to Planetary Regime}",
      journal = {\apj},
         year = 2015,
        month = sep,
       volume = {810},
       number = {2},
          eid = {158},
        pages = {158},
          doi = {10.1088/0004-637X/810/2/158},
archivePrefix = {arXiv},
       eprint = {1508.01767},
 primaryClass = {astro-ph.SR},
       adsurl = {https://ui.adsabs.harvard.edu/abs/2015ApJ...810..158F}
}

@ARTICLE{Franson2024,
       author = {{Franson}, Kyle and {Balmer}, William O. and {Bowler}, Brendan P. and {Pueyo}, Laurent and {Zhou}, Yifan and {Rickman}, Emily and {Zhang}, Zhoujian and {Mukherjee}, Sagnick and {Pearce}, Tim D. and {Bardalez Gagliuffi}, Daniella C. and {Biddle}, Lauren I. and {Brandt}, Timothy D. and {Bowens-Rubin}, Rachel and {Crepp}, Justin R. and {Davidson}, James W. and {Faherty}, Jacqueline and {Ginski}, Christian and {Horch}, Elliott P. and {Morgan}, Marvin and {Morley}, Caroline V. and {Perrin}, Marshall D. and {Sanghi}, Aniket and {Salama}, Ma{\"\i}ssa and {Theissen}, Christopher A. and {Tran}, Quang H. and {Wolf}, Trevor N.},
        title = "{JWST/NIRCam 4─5 {\ensuremath{\mu}}m Imaging of the Giant Planet AF Lep b}",
      journal = {\apjl},
         year = 2024,
        month = oct,
       volume = {974},
       number = {1},
          eid = {L11},
        pages = {L11},
          doi = {10.3847/2041-8213/ad736a},
archivePrefix = {arXiv},
       eprint = {2406.09528},
 primaryClass = {astro-ph.EP},
       adsurl = {https://ui.adsabs.harvard.edu/abs/2024ApJ...974L..11F}
}

@ARTICLE{Gagne2026,
       author = {{Gagn{\'e}}, Jonathan and {Faherty}, Jacqueline K. and {Ruiz Diaz}, Azul and {Coulombe}, Louis-Philippe and {Bickle}, Thomas P. and {Schneider}, Adam C. and {Kirkpatrick}, J. Davy and {Kuchner}, Marc J. and {Meisner}, Aaron M. and {Caselden}, Dan and {Burgasser}, Adam J. and {Casewell}, Sarah and {Honaker}, Easton J. and {Kiwy}, Frank and {Marocco}, Federico and {Bardalez Gagliuffi}, Daniella C. and {Stevnbak Andersen}, Nikolaj and {Ruiz Arroyo}, Lizzeth and {Baller}, Bruce and {Beaulieu}, Paul and {Bell}, John and {Bilsing}, Martin and {Bohling}, Troy K. and {Colin}, Guillaume and {Colombo}, Giovanni and {Deen}, Sam and {Dereveanco}, Alexandru and {Dixon}, Kevin and {Durantini Luca}, Hugo A. and {Flores}, Deiby and {Franck}, Christoph and {Fulvi}, Christopher and {Gallmann}, Michael and {Gantier}, Jean Marc and {Glebov}, Konstantin and {Gramaize}, L{\'e}opold and {Hamlet}, Leslie K. and {Hinckley}, Ken and {Jablonski}, Kevin and {Ja{\l}owiczor}, Peter A. and {Kabatnik}, Martin and {Kasprowitz}, Peter and {Ly}, K and {Martin}, David W. and {Marzak}, Naoufel and {McColgan}, Alexander and {McEwan}, Neil J. and {Michaels}, Marianne N. and {Pendrill}, William and {Perlin}, St{\'e}phane and {Pumphrey}, Ben and {Rabe}, James and {Raway}, Henry and {Robledo}, Walter Ruben and {Roser}, David and {Roy}, Animesh and {Sainio}, Arttu and {Schindler}, Vincent and {Schonau}, Manfred and {Sch{\"u}mann}, J{\"o} rg and {Selg-Mann}, Karl and {Serio}, Andrea and {Smith}, Patrick and {Stenner}, Andres and {Tanner}, Christopher and {Th{\'e}venot}, Melina and {Thakur}, Vinod and {Torres Guerrero}, Mayahuel and {Ventura}, Maurizio and {Voloshin}, Nikita V. and {Walla}, Jim and {W{\c{e}}dracki}, Zbigniew and {Weyandt}, Bailey and {Wilhite}, Breck and {Zitouni}, Spartacus},
        title = "{A SPHEREx Pipeline and Spectral Library for Ultracool Dwarfs}",
      journal = {arXiv e-prints},
         year = 2026,
        month = apr,
          eid = {arXiv:2604.22012},
        pages = {arXiv:2604.22012},
          doi = {10.48550/arXiv.2604.22012},
archivePrefix = {arXiv},
       eprint = {2604.22012},
 primaryClass = {astro-ph.SR},
       adsurl = {https://ui.adsabs.harvard.edu/abs/2026arXiv260422012G}
}

@ARTICLE{Gauza2015,
       author = {{Gauza}, Bartosz and {B{\'e}jar}, Victor J.~S. and {P{\'e}rez-Garrido}, Antonio and {Zapatero Osorio}, Maria Rosa and {Lodieu}, Nicolas and {Rebolo}, Rafael and {Pall{\'e}}, Enric and {Nowak}, Grzegorz},
        title = "{Discovery of a Young Planetary Mass Companion to the Nearby M Dwarf VHS J125601.92-125723.9}",
      journal = {\apj},
         year = 2015,
        month = may,
       volume = {804},
       number = {2},
          eid = {96},
        pages = {96},
          doi = {10.1088/0004-637X/804/2/96},
archivePrefix = {arXiv},
       eprint = {1505.00806},
 primaryClass = {astro-ph.EP},
       adsurl = {https://ui.adsabs.harvard.edu/abs/2015ApJ...804...96G}
}

@ARTICLE{Gibbs2026,
       author = {{Gibbs}, Aidan and {Ruffio}, Jean-Baptiste and {Bidot}, Alexis and {Barman}, Travis S. and {Do {\'O}}, Clarissa R. and {Konopacky}, Quinn M. and {Perrin}, Marshall D. and {Baburaj}, Aneesh and {Dacus}, Beck and {Macintosh}, Bruce and {Madurowicz}, Alex and {Xuan}, Jerry W.},
        title = "{Discovery of an Exterior Third Planet Orbiting {\ensuremath{\beta}} Pictoris}",
      journal = {\apjl},
         year = 2026,
        month = jul,
       volume = {1006},
       number = {1},
          eid = {L11},
        pages = {L11},
          doi = {10.3847/2041-8213/ae801b},
archivePrefix = {arXiv},
       eprint = {2606.23789},
 primaryClass = {astro-ph.EP},
       adsurl = {https://ui.adsabs.harvard.edu/abs/2026ApJ..1006L..11G}
}

@ARTICLE{Golimowski2004,
       author = {{Golimowski}, D.~A. and {Leggett}, S.~K. and {Marley}, M.~S. and {Fan}, X. and {Geballe}, T.~R. and {Knapp}, G.~R. and {Vrba}, F.~J. and {Henden}, A.~A. and {Luginbuhl}, C.~B. and {Guetter}, H.~H. and {Munn}, J.~A. and {Canzian}, B. and {Zheng}, W. and {Tsvetanov}, Z.~I. and {Chiu}, K. and {Glazebrook}, K. and {Hoversten}, E.~A. and {Schneider}, D.~P. and {Brinkmann}, J.},
        title = "{L' and M' Photometry of Ultracool Dwarfs}",
      journal = {\aj},
         year = 2004,
        month = jun,
       volume = {127},
       number = {6},
        pages = {3516-3536},
          doi = {10.1086/420709},
archivePrefix = {arXiv},
       eprint = {astro-ph/0402475},
 primaryClass = {astro-ph},
       adsurl = {https://ui.adsabs.harvard.edu/abs/2004AJ....127.3516G}
}

@ARTICLE{Gorlova2003,
       author = {{Gorlova}, N.~I. and {Meyer}, M.~R. and {Rieke}, G.~H. and {Liebert}, J.},
        title = "{Gravity Indicators in the Near-Infrared Spectra of Brown Dwarfs}",
      journal = {\apj},
         year = 2003,
        month = aug,
       volume = {593},
       number = {2},
        pages = {1074-1092},
          doi = {10.1086/376730},
archivePrefix = {arXiv},
       eprint = {astro-ph/0305147},
 primaryClass = {astro-ph},
       adsurl = {https://ui.adsabs.harvard.edu/abs/2003ApJ...593.1074G}
}

@ARTICLE{Kammerer2024,
       author = {{Kammerer}, Jens and {Lawson}, Kellen and {Perrin}, Marshall D. and {Rebollido}, Isabel and {Stark}, Christopher C. and {Stolker}, Tomas and {Girard}, Julien H. and {Pueyo}, Laurent and {Balmer}, William O. and {Worthen}, Kadin and {Chen}, Christine and {van der Marel}, Roeland P. and {Lewis}, Nikole K. and {Ward-Duong}, Kimberly and {Valenti}, Jeff A. and {Clampin}, Mark and {Mountain}, C. Matt},
        title = "{JWST-TST High Contrast: JWST/NIRCam Observations of the Young Giant Planet {\ensuremath{\beta}} Pic b}",
      journal = {\aj},
         year = 2024,
        month = aug,
       volume = {168},
       number = {2},
          eid = {51},
        pages = {51},
          doi = {10.3847/1538-3881/ad4ffe},
archivePrefix = {arXiv},
       eprint = {2405.18422},
 primaryClass = {astro-ph.EP},
       adsurl = {https://ui.adsabs.harvard.edu/abs/2024AJ....168...51K}
}

@ARTICLE{Kiman2026,
       author = {{Kiman}, Rocio and {Beichman}, Charles A. and {Ruiz Diaz}, Azul and {Faherty}, Jacqueline K. and {Lacy}, Brianna and {Su{\'a}rez}, Genaro and {Marocco}, Federico and {Kirkpatrick}, J. Davy and {Gagn{\'e}}, Jonathan and {Copeland}, Jessica and {Burningham}, Ben and {Whiteford}, Niall and {Rowland}, Melanie J. and {Bardalez Gagliuffi}, Daniella C. and {Vos}, Johanna M. and {Schneider}, Adam C. and {Gonzales}, Eileen C. and {Alejandro Merchan}, Sherelyn and {Rothermich}, Austin and {Smart}, Richard and {Costa}, Edgardo and {Mendez}, Rene A.},
        title = "{The Diversity of Cold Worlds: Age and Characterization of the Exoplanet COCONUTS-2 b}",
      journal = {\aj},
         year = 2026,
        month = feb,
       volume = {171},
       number = {2},
          eid = {60},
        pages = {60},
          doi = {10.3847/1538-3881/ae230f},
archivePrefix = {arXiv},
       eprint = {2511.20923},
 primaryClass = {astro-ph.EP},
       adsurl = {https://ui.adsabs.harvard.edu/abs/2026AJ....171...60K}
}

@ARTICLE{Kirkpatrick2005,
       author = {{Kirkpatrick}, J. Davy},
        title = "{New Spectral Types L and T}",
      journal = {\araa},
         year = 2005,
        month = sep,
       volume = {43},
       number = {1},
        pages = {195-245},
          doi = {10.1146/annurev.astro.42.053102.134017},
       adsurl = {https://ui.adsabs.harvard.edu/abs/2005ARA&A..43..195K}
}

@ARTICLE{Kirkpatrick2006,
       author = {{Kirkpatrick}, J. Davy and {Barman}, Travis S. and {Burgasser}, Adam J. and {McGovern}, Mark R. and {McLean}, Ian S. and {Tinney}, Christopher G. and {Lowrance}, Patrick J.},
        title = "{Discovery of a Very Young Field L Dwarf, 2MASS J01415823-4633574}",
      journal = {\apj},
         year = 2006,
        month = mar,
       volume = {639},
       number = {2},
        pages = {1120-1128},
          doi = {10.1086/499622},
archivePrefix = {arXiv},
       eprint = {astro-ph/0511462},
 primaryClass = {astro-ph},
       adsurl = {https://ui.adsabs.harvard.edu/abs/2006ApJ...639.1120K}
}

@ARTICLE{Kirkpatrick2011,
       author = {{Kirkpatrick}, J. Davy and {Cushing}, Michael C. and {Gelino}, Christopher R. and {Griffith}, Roger L. and {Skrutskie}, Michael F. and {Marsh}, Kenneth A. and {Wright}, Edward L. and {Mainzer}, A. and {Eisenhardt}, Peter R. and {McLean}, Ian S. and {Thompson}, Maggie A. and {Bauer}, James M. and {Benford}, Dominic J. and {Bridge}, Carrie R. and {Lake}, Sean E. and {Petty}, Sara M. and {Stanford}, S.~A. and {Tsai}, Chao-Wei and {Bailey}, Vanessa and {Beichman}, Charles A. and {Bloom}, Joshua S. and {Bochanski}, John J. and {Burgasser}, Adam J. and {Capak}, Peter L. and {Cruz}, Kelle L. and {Hinz}, Philip M. and {Kartaltepe}, Jeyhan S. and {Knox}, Russell P. and {Manohar}, Swarnima and {Masters}, Daniel and {Morales-Calder{\'o}n}, Maria and {Prato}, Lisa A. and {Rodigas}, Timothy J. and {Salvato}, Mara and {Schurr}, Steven D. and {Scoville}, Nicholas Z. and {Simcoe}, Robert A. and {Stapelfeldt}, Karl R. and {Stern}, Daniel and {Stock}, Nathan D. and {Vacca}, William D.},
        title = "{The First Hundred Brown Dwarfs Discovered by the Wide-field Infrared Survey Explorer (WISE)}",
      journal = {\apjs},
         year = 2011,
        month = dec,
       volume = {197},
       number = {2},
          eid = {19},
        pages = {19},
          doi = {10.1088/0067-0049/197/2/19},
archivePrefix = {arXiv},
       eprint = {1108.4677},
 primaryClass = {astro-ph.SR},
       adsurl = {https://ui.adsabs.harvard.edu/abs/2011ApJS..197...19K}
}

@ARTICLE{2021ApJS..253....7K,
       author = {{Kirkpatrick}, J. Davy and {Gelino}, Christopher R. and {Faherty}, Jacqueline K. and {Meisner}, Aaron M. and {Caselden}, Dan and {Schneider}, Adam C. and {Marocco}, Federico and {Cayago}, Alfred J. and {Smart}, R.~L. and {Eisenhardt}, Peter R. and {Kuchner}, Marc J. and {Wright}, Edward L. and {Cushing}, Michael C. and {Allers}, Katelyn N. and {Bardalez Gagliuffi}, Daniella C. and {Burgasser}, Adam J. and {Gagn{\'e}}, Jonathan and {Logsdon}, Sarah E. and {Martin}, Emily C. and {Ingalls}, James G. and {Lowrance}, Patrick J. and {Abrahams}, Ellianna S. and {Aganze}, Christian and {Gerasimov}, Roman and {Gonzales}, Eileen C. and {Hsu}, Chih-Chun and {Kamraj}, Nikita and {Kiman}, Rocio and {Rees}, Jon and {Theissen}, Christopher and {Ammar}, Kareem and {Andersen}, Nikolaj Stevnbak and {Beaulieu}, Paul and {Colin}, Guillaume and {Elachi}, Charles A. and {Goodman}, Samuel J. and {Gramaize}, L{\'e}opold and {Hamlet}, Leslie K. and {Hong}, Justin and {Jonkeren}, Alexander and {Khalil}, Mohammed and {Martin}, David W. and {Pendrill}, William and {Pumphrey}, Benjamin and {Rothermich}, Austin and {Sainio}, Arttu and {Stenner}, Andres and {Tanner}, Christopher and {Th{\'e}venot}, Melina and {Voloshin}, Nikita V. and {Walla}, Jim and {W{\k{e}}dracki}, Zbigniew and {Backyard Worlds: Planet 9 Collaboration}},
        title = "{The Field Substellar Mass Function Based on the Full-sky 20 pc Census of 525 L, T, and Y Dwarfs}",
      journal = {\apjs},
         year = 2021,
        month = mar,
       volume = {253},
       number = {1},
          eid = {7},
        pages = {7},
          doi = {10.3847/1538-4365/abd107},
archivePrefix = {arXiv},
       eprint = {2011.11616},
 primaryClass = {astro-ph.SR},
       adsurl = {https://ui.adsabs.harvard.edu/abs/2021ApJS..253....7K}
}

@ARTICLE{Konopacky2013,
       author = {{Konopacky}, Quinn M. and {Barman}, Travis S. and {Macintosh}, Bruce A. and {Marois}, Christian},
        title = "{Detection of Carbon Monoxide and Water Absorption Lines in an Exoplanet Atmosphere}",
      journal = {Science},
         year = 2013,
        month = mar,
       volume = {339},
       number = {6126},
        pages = {1398-1401},
          doi = {10.1126/science.1232003},
archivePrefix = {arXiv},
       eprint = {1303.3280},
 primaryClass = {astro-ph.EP},
       adsurl = {https://ui.adsabs.harvard.edu/abs/2013Sci...339.1398K}
}

@ARTICLE{Lagrange2009,
       author = {{Lagrange}, A.-M. and {Gratadour}, D. and {Chauvin}, G. and {Fusco}, T. and {Ehrenreich}, D. and {Mouillet}, D. and {Rousset}, G. and {Rouan}, D. and {Allard}, F. and {Gendron}, {\'E}. and {Charton}, J. and {Mugnier}, L. and {Rabou}, P. and {Montri}, J. and {Lacombe}, F.},
        title = "{A probable giant planet imaged in the {\ensuremath{\beta}} Pictoris disk. VLT/NaCo deep L'-band imaging}",
      journal = {\aap},
         year = 2009,
        month = jan,
       volume = {493},
       number = {2},
        pages = {L21-L25},
          doi = {10.1051/0004-6361:200811325},
archivePrefix = {arXiv},
       eprint = {0811.3583},
 primaryClass = {astro-ph},
       adsurl = {https://ui.adsabs.harvard.edu/abs/2009A&A...493L..21L}
}

@ARTICLE{Lagrange2010,
       author = {{Lagrange}, A.-M. and {Bonnefoy}, M. and {Chauvin}, G. and {Apai}, D. and {Ehrenreich}, D. and {Boccaletti}, A. and {Gratadour}, D. and {Rouan}, D. and {Mouillet}, D. and {Lacour}, S. and {Kasper}, M.},
        title = "{A Giant Planet Imaged in the Disk of the Young Star {\ensuremath{\beta}} Pictoris}",
      journal = {Science},
         year = 2010,
        month = jul,
       volume = {329},
       number = {5987},
        pages = {57},
          doi = {10.1126/science.1187187},
archivePrefix = {arXiv},
       eprint = {1006.3314},
 primaryClass = {astro-ph.EP},
       adsurl = {https://ui.adsabs.harvard.edu/abs/2010Sci...329...57L}
}

@ARTICLE{2026AA...709A..56L,
       author = {{Lam}, M.~B. and {Vos}, J.~M. and {Su{\'a}rez}, G. and {Hsu}, C.-C. and {Bickle}, T.~P. and {Faherty}, J. and {Gagn{\'e}}, J. and {Bardalez Gagliuffi}, D. and {Biller}, B. and {Burningham}, B. and {Cruz}, K.~L. and {Morley}, C.~V. and {Luszcz-Cook}, S. and {Lawsky}, S. and {Phillips}, C.~L. and {Rothermich}, A.},
        title = "{Clouds with a silicate lining: Using JWST spectra to probe atmospheric diversity in young AB Dor L dwarfs}",
      journal = {\aap},
         year = 2026,
        month = may,
       volume = {709},
          eid = {A56},
        pages = {A56},
          doi = {10.1051/0004-6361/202558421},
archivePrefix = {arXiv},
       eprint = {2603.24662},
 primaryClass = {astro-ph.SR},
       adsurl = {https://ui.adsabs.harvard.edu/abs/2026A&A...709A..56L}
}

@ARTICLE{Leggett2010,
       author = {{Leggett}, S.~K. and {Burningham}, Ben and {Saumon}, D. and {Marley}, M.~S. and {Warren}, S.~J. and {Smart}, R.~L. and {Jones}, H.~R.~A. and {Lucas}, P.~W. and {Pinfield}, D.~J. and {Tamura}, Motohide},
        title = "{Mid-Infrared Photometry of Cold Brown Dwarfs: Diversity in Age, Mass, and Metallicity}",
      journal = {\apj},
         year = 2010,
        month = feb,
       volume = {710},
       number = {2},
        pages = {1627-1640},
          doi = {10.1088/0004-637X/710/2/1627},
archivePrefix = {arXiv},
       eprint = {1001.0762},
 primaryClass = {astro-ph.SR},
       adsurl = {https://ui.adsabs.harvard.edu/abs/2010ApJ...710.1627L}
}

@ARTICLE{2016ApJ...833...96L,
       author = {{Liu}, Michael C. and {Dupuy}, Trent J. and {Allers}, Katelyn N.},
        title = "{The Hawaii Infrared Parallax Program. II. Young Ultracool Field Dwarfs}",
      journal = {\apj},
         year = 2016,
        month = dec,
       volume = {833},
       number = {1},
          eid = {96},
        pages = {96},
          doi = {10.3847/1538-4357/833/1/96},
archivePrefix = {arXiv},
       eprint = {1612.02426},
 primaryClass = {astro-ph.SR},
       adsurl = {https://ui.adsabs.harvard.edu/abs/2016ApJ...833...96L}
}

@ARTICLE{Looper2007,
       author = {{Looper}, Dagny L. and {Kirkpatrick}, J. Davy and {Burgasser}, Adam J.},
        title = "{Discovery of 11 New T Dwarfs in the Two Micron All Sky Survey, Including a Possible L/T Transition Binary}",
      journal = {\aj},
         year = 2007,
        month = sep,
       volume = {134},
       number = {3},
        pages = {1162-1182},
          doi = {10.1086/520645},
archivePrefix = {arXiv},
       eprint = {0706.1601},
 primaryClass = {astro-ph},
       adsurl = {https://ui.adsabs.harvard.edu/abs/2007AJ....134.1162L}
}

@ARTICLE{Looper2008,
       author = {{Looper}, Dagny L. and {Kirkpatrick}, J. Davy and {Cutri}, Roc M. and {Barman}, Travis and {Burgasser}, Adam J. and {Cushing}, Michael C. and {Roellig}, Thomas and {McGovern}, Mark R. and {McLean}, Ian S. and {Rice}, Emily and {Swift}, Brandon J. and {Schurr}, Steven D.},
        title = "{Discovery of Two Nearby Peculiar L Dwarfs from the 2MASS Proper-Motion Survey: Young or Metal-Rich?}",
      journal = {\apj},
         year = 2008,
        month = oct,
       volume = {686},
       number = {1},
        pages = {528-541},
          doi = {10.1086/591025},
archivePrefix = {arXiv},
       eprint = {0806.1059},
 primaryClass = {astro-ph},
       adsurl = {https://ui.adsabs.harvard.edu/abs/2008ApJ...686..528L}
}

@ARTICLE{Lucas2001,
       author = {{Lucas}, P.~W. and {Roche}, P.~F. and {Allard}, France and {Hauschildt}, P.~H.},
        title = "{Infrared spectroscopy of substellar objects in Orion}",
      journal = {\mnras},
         year = 2001,
        month = sep,
       volume = {326},
       number = {2},
        pages = {695-721},
          doi = {10.1046/j.1365-8711.2001.04666.x},
archivePrefix = {arXiv},
       eprint = {astro-ph/0105154},
 primaryClass = {astro-ph},
       adsurl = {https://ui.adsabs.harvard.edu/abs/2001MNRAS.326..695L}
}

@ARTICLE{Lueber2022,
       author = {{Lueber}, Anna and {Kitzmann}, Daniel and {Bowler}, Brendan P. and {Burgasser}, Adam J. and {Heng}, Kevin},
        title = "{Retrieval Study of Brown Dwarfs across the L-T Sequence}",
      journal = {\apj},
         year = 2022,
        month = may,
       volume = {930},
       number = {2},
          eid = {136},
        pages = {136},
          doi = {10.3847/1538-4357/ac63b9},
archivePrefix = {arXiv},
       eprint = {2204.01330},
 primaryClass = {astro-ph.EP},
       adsurl = {https://ui.adsabs.harvard.edu/abs/2022ApJ...930..136L}
}

@ARTICLE{Manjavacas2020,
       author = {{Manjavacas}, E. and {Lodieu}, N. and {B{\'e}jar}, V.~J.~S. and {Zapatero-Osorio}, M.~R. and {Boudreault}, S. and {Bonnefoy}, M.},
        title = "{Spectral library of age-benchmark low-mass stars and brown dwmarfs}",
      journal = {\mnras},
         year = 2020,
        month = feb,
       volume = {491},
       number = {4},
        pages = {5925-5950},
          doi = {10.1093/mnras/stz3441},
archivePrefix = {arXiv},
       eprint = {1912.02806},
 primaryClass = {astro-ph.SR},
       adsurl = {https://ui.adsabs.harvard.edu/abs/2020MNRAS.491.5925M}
}

@ARTICLE{Manjavacas2024,
       author = {{Manjavacas}, Elena and {Tremblin}, Pascal and {Birkmann}, Stephan and {Valenti}, Jeff and {Alves de Oliveira}, Catarina and {Beck}, Tracy L. and {Giardino}, G. and {L{\"u}tzgendorf}, N. and {Rauscher}, B.~J. and {Sirianni}, M.},
        title = "{Medium-resolution 0.97─5.3 {\ensuremath{\mu}}m Spectra of Very Young Benchmark Brown Dwarfs with NIRSpec on Board the James Webb Space Telescope}",
      journal = {\aj},
         year = 2024,
        month = apr,
       volume = {167},
       number = {4},
          eid = {168},
        pages = {168},
          doi = {10.3847/1538-3881/ad2938},
archivePrefix = {arXiv},
       eprint = {2402.04230},
 primaryClass = {astro-ph.SR},
       adsurl = {https://ui.adsabs.harvard.edu/abs/2024AJ....167..168M}
}

@ARTICLE{Marois2008,
       author = {{Marois}, Christian and {Macintosh}, Bruce and {Barman}, Travis and {Zuckerman}, B. and {Song}, Inseok and {Patience}, Jennifer and {Lafreni{\`e}re}, David and {Doyon}, Ren{\'e}},
        title = "{Direct Imaging of Multiple Planets Orbiting the Star HR 8799}",
      journal = {Science},
         year = 2008,
        month = nov,
       volume = {322},
       number = {5906},
        pages = {1348},
          doi = {10.1126/science.1166585},
archivePrefix = {arXiv},
       eprint = {0811.2606},
 primaryClass = {astro-ph},
       adsurl = {https://ui.adsabs.harvard.edu/abs/2008Sci...322.1348M}
}

@ARTICLE{Marocco2014,
       author = {{Marocco}, F. and {Day-Jones}, A.~C. and {Lucas}, P.~W. and {Jones}, H.~R.~A. and {Smart}, R.~L. and {Zhang}, Z.~H. and {Gomes}, J.~I. and {Burningham}, B. and {Pinfield}, D.~J. and {Raddi}, R. and {Smith}, L.},
        title = "{The extremely red L dwarf ULAS J222711-004547 - dominated by dust}",
      journal = {\mnras},
         year = 2014,
        month = mar,
       volume = {439},
       number = {1},
        pages = {372-386},
          doi = {10.1093/mnras/stt2463},
archivePrefix = {arXiv},
       eprint = {1401.0420},
 primaryClass = {astro-ph.SR},
       adsurl = {https://ui.adsabs.harvard.edu/abs/2014MNRAS.439..372M}
}

@ARTICLE{Marois2010,
       author = {{Marois}, Christian and {Zuckerman}, B. and {Konopacky}, Quinn M. and {Macintosh}, Bruce and {Barman}, Travis},
        title = "{Images of a fourth planet orbiting HR 8799}",
      journal = {\nat},
         year = 2010,
        month = dec,
       volume = {468},
       number = {7327},
        pages = {1080-1083},
          doi = {10.1038/nature09684},
archivePrefix = {arXiv},
       eprint = {1011.4918},
 primaryClass = {astro-ph.EP},
       adsurl = {https://ui.adsabs.harvard.edu/abs/2010Natur.468.1080M}
}

@ARTICLE{Miles2018,
       author = {{Miles}, Brittany E. and {Skemer}, Andrew J. and {Barman}, Travis S. and {Allers}, Katelyn N. and {Stone}, Jordan M.},
        title = "{Methane in Analogs of Young Directly Imaged Exoplanets}",
      journal = {\apj},
         year = 2018,
        month = dec,
       volume = {869},
       number = {1},
          eid = {18},
        pages = {18},
          doi = {10.3847/1538-4357/aae6cd},
archivePrefix = {arXiv},
       eprint = {1810.04684},
 primaryClass = {astro-ph.EP},
       adsurl = {https://ui.adsabs.harvard.edu/abs/2018ApJ...869...18M}
}

@ARTICLE{Miles2023,
       author = {{Miles}, Brittany E. and {Biller}, Beth A. and {Patapis}, Polychronis and {Worthen}, Kadin and {Rickman}, Emily and {Hoch}, Kielan K.~W. and {Skemer}, Andrew and {Perrin}, Marshall D. and {Whiteford}, Niall and {Chen}, Christine H. and {Sargent}, B. and {Mukherjee}, Sagnick and {Morley}, Caroline V. and {Moran}, Sarah E. and {Bonnefoy}, Mickael and {Petrus}, Simon and {Carter}, Aarynn L. and {Choquet}, Elodie and {Hinkley}, Sasha and {Ward-Duong}, Kimberly and {Leisenring}, Jarron M. and {Millar-Blanchaer}, Maxwell A. and {Pueyo}, Laurent and {Ray}, Shrishmoy and {Sallum}, Steph and {Stapelfeldt}, Karl R. and {Stone}, Jordan M. and {Wang}, Jason J. and {Absil}, Olivier and {Balmer}, William O. and {Boccaletti}, Anthony and {Bonavita}, Mariangela and {Booth}, Mark and {Bowler}, Brendan P. and {Chauvin}, Gael and {Christiaens}, Valentin and {Currie}, Thayne and {Danielski}, Camilla and {Fortney}, Jonathan J. and {Girard}, Julien H. and {Grady}, Carol A. and {Greenbaum}, Alexandra Z. and {Henning}, Thomas and {Hines}, Dean C. and {Janson}, Markus and {Kalas}, Paul and {Kammerer}, Jens and {Kennedy}, Grant M. and {Kenworthy}, Matthew A. and {Kervella}, Pierre and {Lagage}, Pierre-Olivier and {Lew}, Ben W.~P. and {Liu}, Michael C. and {Macintosh}, Bruce and {Marino}, Sebastian and {Marley}, Mark S. and {Marois}, Christian and {Matthews}, Elisabeth C. and {Matthews}, Brenda C. and {Mawet}, Dimitri and {McElwain}, Michael W. and {Metchev}, Stanimir and {Meyer}, Michael R. and {Molliere}, Paul and {Pantin}, Eric and {Quirrenbach}, Andreas and {Rebollido}, Isabel and {Ren}, Bin B. and {Schneider}, Glenn and {Vasist}, Malavika and {Wyatt}, Mark C. and {Zhou}, Yifan and {Briesemeister}, Zackery W. and {Bryan}, Marta L. and {Calissendorff}, Per and {Cantalloube}, Faustine and {Cugno}, Gabriele and {De Furio}, Matthew and {Dupuy}, Trent J. and {Factor}, Samuel M. and {Faherty}, Jacqueline K. and {Fitzgerald}, Michael P. and {Franson}, Kyle and {Gonzales}, Eileen C. and {Hood}, Callie E. and {Howe}, Alex R. and {Kraus}, Adam L. and {Kuzuhara}, Masayuki and {Lagrange}, Anne-Marie and {Lawson}, Kellen and {Lazzoni}, Cecilia and {Liu}, Pengyu and {Llop-Sayson}, Jorge and {Lloyd}, James P. and {Martinez}, Raquel A. and {Mazoyer}, Johan and {Quanz}, Sascha P. and {Redai}, Jea Adams and {Samland}, Matthias and {Schlieder}, Joshua E. and {Tamura}, Motohide and {Tan}, Xianyu and {Uyama}, Taichi and {Vigan}, Arthur and {Vos}, Johanna M. and {Wagner}, Kevin and {Wolff}, Schuyler G. and {Ygouf}, Marie and {Zhang}, Xi and {Zhang}, Keming and {Zhang}, Zhoujian},
        title = "{The JWST Early-release Science Program for Direct Observations of Exoplanetary Systems II: A 1 to 20 {\ensuremath{\mu}}m Spectrum of the Planetary-mass Companion VHS 1256-1257 b}",
      journal = {\apjl},
         year = 2023,
        month = mar,
       volume = {946},
       number = {1},
          eid = {L6},
        pages = {L6},
          doi = {10.3847/2041-8213/acb04a},
archivePrefix = {arXiv},
       eprint = {2209.00620},
 primaryClass = {astro-ph.EP},
       adsurl = {https://ui.adsabs.harvard.edu/abs/2023ApJ...946L...6M}
}

@ARTICLE{Morley2018,
       author = {{Morley}, Caroline V. and {Skemer}, Andrew J. and {Allers}, Katelyn N. and {Marley}, Mark. S. and {Faherty}, Jacqueline K. and {Visscher}, Channon and {Beiler}, Samuel A. and {Miles}, Brittany E. and {Lupu}, Roxana and {Freedman}, Richard S. and {Fortney}, Jonathan J. and {Geballe}, Thomas R. and {Bjoraker}, Gordon L.},
        title = "{An L Band Spectrum of the Coldest Brown Dwarf}",
      journal = {\apj},
         year = 2018,
        month = may,
       volume = {858},
       number = {2},
          eid = {97},
        pages = {97},
          doi = {10.3847/1538-4357/aabe8b},
archivePrefix = {arXiv},
       eprint = {1804.07771},
 primaryClass = {astro-ph.EP},
       adsurl = {https://ui.adsabs.harvard.edu/abs/2018ApJ...858...97M}
}

@ARTICLE{Piscarreta2024,
       author = {{Piscarreta}, L. and {Mu{\v{z}}i{\'c}}, K. and {Almendros-Abad}, V. and {Scholz}, A.},
        title = "{Spectral characterization of young LT dwarfs}",
      journal = {\aap},
         year = 2024,
        month = jun,
       volume = {686},
          eid = {A37},
        pages = {A37},
          doi = {10.1051/0004-6361/202347327},
archivePrefix = {arXiv},
       eprint = {2402.16802},
 primaryClass = {astro-ph.SR},
       adsurl = {https://ui.adsabs.harvard.edu/abs/2024A&A...686A..37P}
}

@ARTICLE{Rayner2003,
       author = {{Rayner}, J.~T. and {Toomey}, D.~W. and {Onaka}, P.~M. and {Denault}, A.~J. and {Stahlberger}, W.~E. and {Vacca}, W.~D. and {Cushing}, M.~C. and {Wang}, S.},
        title = "{SpeX: A Medium-Resolution 0.8-5.5 Micron Spectrograph and Imager for the NASA Infrared Telescope Facility}",
      journal = {\pasp},
         year = 2003,
        month = mar,
       volume = {115},
       number = {805},
        pages = {362-382},
          doi = {10.1086/367745},
       adsurl = {https://ui.adsabs.harvard.edu/abs/2003PASP..115..362R}
}

@ARTICLE{Reid2006,
       author = {{Reid}, I. Neill and {Lewitus}, E. and {Burgasser}, Adam J. and {Cruz}, K.~L.},
        title = "{2MASS J22521073-1730134: A Resolved L/T Binary at 14 Parsecs}",
      journal = {\apj},
         year = 2006,
        month = mar,
       volume = {639},
       number = {2},
        pages = {1114-1119},
          doi = {10.1086/499484},
       adsurl = {https://ui.adsabs.harvard.edu/abs/2006ApJ...639.1114R}
}

@ARTICLE{Ruffio2026,
       author = {{Ruffio}, Jean-Baptiste and {Xuan}, Jerry W. and {Chachan}, Yayaati and {Kesseli}, Aurora and {Lee}, Eve J. and {Beichman}, Charles and {Hodapp}, Klaus and {Balmer}, William O. and {Konopacky}, Quinn and {Perrin}, Marshall D. and {Mawet}, Dimitri and {Knutson}, Heather A. and {Bryden}, Geoffrey and {Greene}, Thomas P. and {Johnstone}, Doug and {Leisenring}, Jarron and {Meyer}, Michael and {Ygouf}, Marie},
        title = "{Jupiter-like uniform metal enrichment in a system of multiple giant exoplanets}",
      journal = {Nature Astronomy},
         year = 2026,
        month = apr,
       volume = {10},
        pages = {511-521},
          doi = {10.1038/s41550-026-02783-z},
archivePrefix = {arXiv},
       eprint = {2601.08227},
 primaryClass = {astro-ph.EP},
       adsurl = {https://ui.adsabs.harvard.edu/abs/2026NatAs..10..511R}
}

@ARTICLE{Rustamkulov2026,
       author = {{Rustamkulov}, Zafar and {Kirkpatrick}, J. and {Akeson}, Rachel and {Werner}, Michael W and {Ashby}, Matthew and {Chang}, Tzu-Ching and {Chen}, Shuang-Shuang and {Cooray}, Asantha and {Crill}, Brendan and {Dore}, Olivier and {Dowell}, C. and {Faisst}, Andreas and {Hui}, Howard and {Jeong}, Woong-Seob and {Kang}, Miju and {Korngut}, Phil and {Lisse}, Carey and {Masters}, Daniel and {Melnick}, Gary and {Nguyen}, Chi and {Paladini}, Roberta and {Tolls}, Volker and {Yang}, Yujin and {Zemcov}, Michael},
        title = "{SPHEREx 0.75 to 5 $μ$m Spectra for a Sequence of Nearby Brown Dwarfs}",
      journal = {arXiv e-prints},
         year = 2026,
        month = jul,
          eid = {arXiv:2607.00543},
        pages = {arXiv:2607.00543},
          doi = {10.48550/arXiv.2607.00543},
archivePrefix = {arXiv},
       eprint = {2607.00543},
 primaryClass = {astro-ph.SR},
       adsurl = {https://ui.adsabs.harvard.edu/abs/2026arXiv260700543R}
}

@ARTICLE{2017AJ....153..196S,
       author = {{Schneider}, Adam C. and {Windsor}, James and {Cushing}, Michael C. and {Kirkpatrick}, J. Davy and {Shkolnik}, Evgenya L.},
        title = "{A 2MASS/AllWISE Search for Extremely Red L Dwarfs: The Discovery of Several Likely L Type Members of {\ensuremath{\beta}} Pic, AB Dor, Tuc-Hor, Argus, and the Hyades}",
      journal = {\aj},
         year = 2017,
        month = apr,
       volume = {153},
       number = {4},
          eid = {196},
        pages = {196},
          doi = {10.3847/1538-3881/aa6624},
archivePrefix = {arXiv},
       eprint = {1703.03774},
 primaryClass = {astro-ph.SR},
       adsurl = {https://ui.adsabs.harvard.edu/abs/2017AJ....153..196S}
}

@misc{spherex,
  doi = {10.26131/IRSA652},
  url = {https://catcopy.ipac.caltech.edu/dois/doi.php?id=10.26131/IRSA652},
  author = {{SPHEREx Team}},
  title = {SPHEREx Quick Release Spectral Images - QR2},
  publisher = {IPAC},
  year = {2025}
}

@PHDTHESIS{2016PhDT.......189A,
       author = {{Aller}, Kimberly Mei},
        title = "{Finding the elusive substellar members of young moving groups}",
       school = {University of Hawaii, Manoa},
         year = 2016,
        month = jan,
       adsurl = {https://ui.adsabs.harvard.edu/abs/2016PhDT.......189A}
}
\bibliographystyle{aasjournalv7}



\end{document}